\documentclass{elsart}
\usepackage{graphicx}
\newcommand{\hs}{\hspace*{0.5cm}}
\newcommand{\vs}{\vspace*{0.5cm}}
\newcommand{\be}{\begin{equation}}
\newcommand{\ee}{\end{equation}}
\newcommand{\bea}{\begin{eqnarray}}
\newcommand{\eea}{\end{eqnarray}}
\newcommand{\nn}{\nonumber}
\newcommand{\crn}{\nonumber \\}

\newcommand{\al}{\alpha}
\newcommand{\la}{\lambda}
\newcommand{\bet}{\beta}
\newcommand{\ga}{\gamma}
\newcommand{\va}{\varphi}

\newcommand{\pa}{\partial}
\newcommand{\fr}{\frac}
\newcommand{\bc}{\begin{center}}
\newcommand{\ec}{\end{center}}
\newcommand{\Ga}{\Gamma}

\newcommand {\ba}{\begin{array}}
\newcommand {\ea}{\end{array}}
\newcommand{\ben}{\begin{enumerate}}
\newcommand{\een}{\end{enumerate}}

\begin{document}
\begin{frontmatter}
\title{Higgs phenomenology of supersymmetric economical 3-3-1 model} \vspace{0.1cm}

P. V. Dong, D. T. Huong, N. T. Thuy, H. N. Long

\vs

{\it Institute of Physics, VAST, P.O. Box 429, Bo Ho, Hanoi 10000,
Vietnam}

\begin{abstract}
We explore the Higgs sector in the supersymmetric economical 3-3-1
model and find new features in this sector. The charged Higgs sector
is revised i.e., in difference of the previous work, the exact
eigenvalues and states are obtained without any approximation. In
this model, there are three Higgs bosons having masses equal to that
of the gauge bosons---the $W$ and extra $X$ and $Y$. There is one
scalar boson with mass of 91.4 GeV, which is closed to the $Z$ boson
mass and in good agreement with present limit: 89.8 GeV at 95\% CL.
The condition of eliminating for charged scalar tachyon leads to
splitting of VEV at the first symmetry breaking, namely, $w \simeq
w^\prime$. The interactions among the standard model gauge bosons
and scalar fields  in the framework of the supersymmetric economical
3-3-1 model are presented. From these couplings, at some limit,
almost scalar Higgs fields can be recognized in accordance with the
standard model. The hadronic cross section for production of the
bilepton charged Higgs boson at the CERN LHC in the effective vector
boson approximation is calculated. Numerical evaluation shows that
the cross section can exceed 35.8 $fb$.

\end{abstract}
\end{frontmatter}

\section{\label{intro}Introduction}
Recent neutrino  experimental results~\cite{superk,kam,sno}
establish the fact that neutrinos have masses and the standard
model (SM) must be extended. The generation of neutrino masses is
thus an important issue in any realistic extension of the SM. In
general, the values of these masses which are of the order of, or
less than, 1 eV needed to explain all neutrino oscillation data
are not enough to put strong constraints on model building. It
means that several models can induce neutrino masses and mixing
compatible with experimental data. In such cases it is more useful
to consider in any particular model motivation other than that can
explain neutrino masses. In addition, although the SM is
exceedingly successful in describing charged leptons, quarks and
their interactions, it is not considered as the ultimate theory
since neither the fundamental parameters, masses and couplings,
nor the symmetry pattern are predicted. These elements are merely
built into the model. Likewise, the spontaneous electroweak
symmetry breaking is simply parametrized by a single Higgs doublet
field.

The embedding of the model into a more general framework is
therefore expected. If the Higgs boson is light, the SM can
naturally be embedded in a grand unified theory. The large energy
gap between the low electroweak scale and the high grand unification
scale can be stabilized by a supersymmetry (SUSY) transforming
bosons into fermions and vice versa \cite{susy}. The existence of
such a non-trivial extension is highly constrained by theoretical
principles and actually provides the link between the experimentally
explored interactions at electroweak energy scales and physics at
scales close to the Planck scale $M_{pl} \approx 10^{19}$ GeV where
gravity is important.  One of the intriguing features of the
supersymmetric models is that the Higgs spectrum is quite
constrained. This statement is consolidated by our analysis below.

On the other hand, the possibility of a gauge symmetry based on
$\mathrm{SU}(3)_{C} \otimes \mathrm{SU}(3)_{L} \otimes
\mathrm{U}(1)_{X}$ (3-3-1)~\cite{ppf,flt,331rh} is particularly
interesting, because it explains some fundamental questions that are
eluded in the SM. The main motivations to study this kind of model
are:
\begin{enumerate}
\item The family number must be multiple of three;
\item It solves the strong CP problem; \item It is the simplest
model that includes bileptons of both types: scalar and vectors
ones; \item The model has several sources of CP violation. \item
The explaination of electric charge quantization~\cite{chargequan}
\end{enumerate}

In one of 3-3-1 models \cite{331rh}, the anomaly-free particle
content is given by \bea L_{aL} &=& \left(\nu_{a}, l_{a},
\nu_{a}^c \right)^T_L \sim (1, 3, -1/3),\hs l_{aR}\sim (1,1,
-1),\hs a = 1, 2, 3, \crn Q_{1L}&=&\left( u_{1},  d_{1}, u'
\right)^T_L\sim \left(3,3,1/3\right),\crn Q_{\al
L}&=&\left(d_{\al},  -u_{\al}, d'_{\al} \right)^T_L\sim
(3,3^*,0),\hs \al=2,3,\crn u_{i R}&\sim&\left(3,1,2/3\right),\hs
d_{i R} \sim \left(3,1,-1/3\right),\hs i=1,2,3, \crn u'_{R}&\sim&
\left(3,1,2/3\right),\hs d'_{\al R} \sim \left(3,1,-1/3\right),
\nn\eea where the values in the parentheses denote quantum numbers
based on the
$\left(\mbox{SU}(3)_C,\mbox{SU}(3)_L,\mbox{U}(1)_X\right)$
symmetry. The exotic quarks $u'$ and $d'_\al$ take the same
electric charges as of the usual quarks, i.e., $q_{u'}=2/3$,
$q_{d'_\al}=-1/3$. The spontaneous symmetry breaking is achieved
by two Higgs scalar triplets only \be \chi =\left(\chi^0_1,
\chi^-, \chi^0_2 \right)^T \sim \left(1,3,-1/3\right),\hs
\rho=\left(\rho^+_1, \rho^0, \rho^+_2\right)^T \sim
\left(1,3,2/3\right)\ee with all the neutral components
$\chi^0_{1}$, $\chi^0_2$ and $\rho^0$ developing the vacuum
expectation values (VEVs). Such a scalar sector is minimal,
therefore it has been called the economical 3-3-1
model~\cite{ponce,haihiggs}.

In a series of papers, we have developed and proved that this
version is consistent, realistic and very rich in physics. Let us
remind some steps in the development: The general Higgs sector is
very simple and consists of three physical scalars (two neutral
and one charged) and eight Goldstone bosons---the needed number
for massive gauge ones~\cite{higgseconom}. In
Refs.\cite{dlhh,dls1}, we have  shown that the model under the
consideration is realistic, by the mean that, at the  one-loop
level, all fermions gain consistent masses. It was shown
that~\cite{higgseconom} the economical 3-3-1 model does not
furnish any candidate for self-interaction dark matter. This
directly relates to the scalar sector in which a significant
number of fields and couplings is reduced. With a larger field
content in order to provide candidates for dark matter, the
supersymmetric version of the economical 3-3-1 model has already
been constructed in Ref. \cite{susyec}.

It is well known that the electroweak symmetry breaking  in the SM
is achieved via the Higgs mechanism. In the Glashow-Weinberg-Salam
model there is a single complex Higgs doublet, where the Higgs
boson $h$ is the physical neutral Higgs scalar which is the only
remaining part of this doublet after spontaneous symmetry
breaking. In the extended  models there are additional charged and
neutral scalar Higgs particles. The prospects for Higgs coupling
measurements at the CERN LHC have recently been analyzed in detail
in Ref.~\cite{logan}. The experimental detection of the $h$ will
be great triumph of the SM of electroweak interactions and will
mark new stage in high energy physics.

In extended Higgs models, which would be deduced in the low energy
effective theory of new physics models, additional Higgs bosons
like charged and CP-odd scalar bosons are predicted. Unlike the
spectrum of squarks, sleptons and gauginos, which are determined
by many parameters, the Higgs spectrum is quite constrained.
Phenomenology of these extra scalar bosons strongly depends on the
characteristics of each new physics model. By measuring their
properties like masses, widths, production rates and decay
branching ratios, the outline of physics beyond the electroweak
scale can be experimentally determined. In the model under
consideration, at the tree level, the mass lightest Higgs is the
mass of the $W$ boson. This is in agreement with the current
experimental limit.

The interesting feature compared with other 3-3-1 models is the
Higgs physics. In the 3-3-1 models, the general Higgs sector is very
complicated~\cite{changlong,scalrh} and this prevents the models'
predicability. Thus, the Higgs sector of the supersymmetric version
of the 3-3-1 models are intricate too \cite{hrl,shiggs}. The Higgs
sector of an supersymmetric version of the economical 3-3-1 model is
not so complicated and  its eigenvalues and states can be found
exactly without any approximation. The scalar sector of the
supersymmetric economical 3-3-1 model is a subject of the present
study. As shown, by couplings of the scalar fields with the ordinary
gauge bosons such as the photon, the $W$ and the neutral $Z$ gauge
bosons, we are able to identify full content of the Higgs sector in
the SM including the neutral $h$ and the Goldstone bosons eaten by
their associated massive gauge ones. Almost interactions among
Higgs-gauge bosons in the standard model are {\it recovered}.

The aim of this work is to explore more features of the
supersymmetric version of the economical 3-3-1 model through  the
Higgs-gauge boson interactions. In scalar sector of the model, there
exists the singly-charged boson $\zeta_4^\pm$, which is a subject of
intensive current studies (see, for example, Ref.~\cite{kame,roy}).
The trilinear coupling $ZW^\pm \zeta^\mp_4$ which differs, at the
tree level, from zero  only in the models with Higgs triplets, plays
a special role on study phenomenology of these exotic
representations. We shall pay particular interest on this boson.

The outline of this paper is as follows.  Sec.\ref{model} is
devoted to a brief review of the model. The scalar fields and mass
spectrum is revisited in Sec.\ref{sec:massspectrum} and their
couplings with the ordinary gauge bosons are given in
Sec.\ref{susyhg}. Production of the heavy singly charged Higgs
boson  $\zeta^\pm_4$ at the CERN LHC are calculated in
Sec.\ref{chargeds}. We  outline our main results in the last
section - Sec.\ref{concl}.

\section{A review of the model}
\label{model} In this section we first recapitulate the basic
elements  of the model \cite{susyec}.

\subsection{\label{parcontent}Particle content }

 The superfield content in this paper is defined in a standard way as
follows \be \widehat{F}= (\widetilde{F}, F),\hs \widehat{S} = (S,
\widetilde{S}),\hs \widehat{V}= (\lambda,V), \ee where the
components $F$, $S$ and $V$ stand for the fermion, scalar and
vector fields while their superpartners are denoted as
$\widetilde{F}$, $\widetilde{S}$ and $\lambda$, respectively
\cite{susy,s331r}.

The superfields for the leptons under the 3-3-1 gauge group
transform as
\begin{equation}
\widehat{L}_{a L}=\left(\widehat{\nu}_{a}, \widehat{l}_{a},
\widehat{\nu}^c_{a}\right)^T_{L} \sim (1,3,-1/3),\hs
  \widehat {l}^{c}_{a L} \sim (1,1,1),\label{l2}
\end{equation} where $\widehat{\nu}^c_L=(\widehat{\nu}_R)^c$ and $a=1,2,3$
is a generation index.

The superfields for the left-handed quarks of the first generation
are in triplets \be \widehat Q_{1L}= \left(\widehat { u}_1,\
                        \widehat {d}_1,\
                        \widehat {u}^\prime
 \right)^T_L \sim (3,3,1/3),\label{quarks3}\ee
where the right-handed singlet counterparts are given by\be
\widehat {u}^{c}_{1L},\ \widehat { u}^{ \prime c}_{L} \sim
(3^*,1,-2/3),\hs \widehat {d}^{c}_{1L} \sim (3^*,1,1/3 ).
\label{l5} \ee Conversely, the superfields for the last two
generations transform as antitriplets
\begin{equation}
\begin{array}{ccc}
 \widehat{Q}_{\alpha L} = \left(\widehat{d}_{\alpha}, - \widehat{u}_{\alpha},
 \widehat{d^\prime}_{\alpha}\right)^T_{L} \sim (3,3^*,0), \hs \al=2,3, \label{l3}
\end{array}
\end{equation}
where the right-handed counterparts are in singlets
\begin{equation}
\widehat{u}^{c}_{\alpha L} \sim \left(3^*,1,-2/3 \right),\hs
\widehat{d}^{c}_{\alpha L},\ \widehat{d}^{\prime c}_{\alpha L}
\sim \left(3^*,1,1/3 \right). \label{l4}
\end{equation}

The primes superscript on usual quark types ($u'$ with the
electric charge $q_{u'}=2/3$ and $d'$ with $q_{d'}=-1/3$) indicate
that those quarks are exotic ones. The mentioned fermion content,
which belongs to that of the 3-3-1 model with right-handed
neutrinos \cite{331rh,haihiggs} is, of course,  free from anomaly.

The two superfields $\widehat{\chi}$ and $\widehat {\rho} $ are at
least introduced to span the scalar sector of the economical 3-3-1
model \cite{higgseconom}: \bea \widehat{\chi}&=& \left (
\widehat{\chi}^0_1, \widehat{\chi}^-, \widehat{\chi}^0_2
\right)^T\sim (1,3,-1/3), \label{l7}\\
\widehat{\rho}&=& \left (\widehat{\rho}^+_1, \widehat{\rho}^0,
\widehat{\rho}^+_2\right)^T \sim  (1,3,2/3). \label{l8} \eea To
cancel the chiral anomalies of Higgsino sector, the two extra
superfields $\widehat{\chi}^\prime$ and $\widehat {\rho}^\prime $
must be added as follows \bea \widehat{\chi}^\prime&=& \left
(\widehat{\chi}^{\prime 0}_1, \widehat{\chi}^{\prime
+},\widehat{\chi}^{\prime 0}_2 \right)^T\sim ( 1,3^*,1/3),
\label{l9}\\
\widehat{\rho}^\prime &=& \left (\widehat{\rho}^{\prime -}_1,
  \widehat{\rho}^{\prime 0},  \widehat{\rho}^{\prime -}_2
\right)^T\sim (1,3^*,-2/3). \label{l10} \eea

In this model, the $ \mathrm{SU}(3)_L \otimes \mathrm{U}(1)_X$
gauge group is broken via two steps:
 \be \mathrm{SU}(3)_L \otimes
\mathrm{U}(1)_X \stackrel{w,w'}{\longrightarrow}\ \mathrm{SU}(2)_L
\otimes \mathrm{U}(1)_Y\stackrel{v,v',u,u'}{\longrightarrow}
\mathrm{U}(1)_{Q},\label{stages}\ee where the VEVs are defined by
\bea
 \sqrt{2} \langle\chi\rangle^T &=& \left(u, 0, w\right), \hs \sqrt{2}
 \langle\chi^\prime\rangle^T = \left(u^\prime,  0,
 w^\prime\right),\\
\sqrt{2}  \langle\rho\rangle^T &=& \left( 0, v, 0 \right), \hs
\sqrt{2} \langle\rho^\prime\rangle^T = \left( 0, v^\prime,  0
\right).\eea The VEVs $w$ and $w^\prime$ are responsible for the
first step of the symmetry breaking while $u,\ u^\prime$ and $v,\
v^\prime$ are for the second one. Therefore, they have to satisfy
the constraints:
 \be
 u,\ u^\prime,\ v,\ v^\prime
\ll w,\ w^\prime. \label{contraint}\ee

The vector superfields $\widehat{V}_c$, $\widehat{V}$ and
$\widehat{V}^\prime$ containing the usual gauge bosons are,
respectively, associated with the $\mathrm{SU}(3)_C$,
$\mathrm{SU}(3)_L$ and $\mathrm{U}(1)_X $ group factors. The colour
and flavour vector superfields have expansions in the Gell-Mann
matrix bases $T^a=\lambda^a/2$ $(a=1,2,...,8)$ as follows\bea
\widehat{V}_c &=& \fr{1}{2}\lambda^a \widehat{V}_{ca},\hs
\widehat{\overline{V}}_c=-\fr{1}{2}\lambda^{a*} \widehat{V}_{ca};\hs
\widehat{V} = \fr{1}{2}\lambda^a \widehat{V}_{a},\hs
\widehat{\overline{V}}=-\fr{1}{2}\lambda^{a*} \widehat{V}_{a},\eea
where an overbar $^-$ indicates complex conjugation. For the vector
superfield associated with $\mathrm{U}(1)_X$, we normalize as
follows \be X \hat{V}'= (XT^9) \hat{B}, \hs
T^9\equiv\fr{1}{\sqrt{6}}\mathrm{diag}(1,1,1).\ee In the following,
we are denoting the gluons by $g^a$ and their respective gluino
partners by $\lambda^a_{c}$, with $a=1, \ldots,8$. In the
electroweak sector, $V^a$ and $B$ stand for the $\mathrm{SU}(3)_{L}$
and $\mathrm{U}(1)_{X}$ gauge bosons with their gaugino partners
$\lambda^a_{V}$ and $\lambda_{B}$, respectively.

The supersymmetric model possesses a full Lagrangian of the form
$\mathcal{L}_{susy}+\mathcal{L}_{soft}$, where the first term is
supersymmetric part, whereas the last term breaks explicitly the
supersymmetry. We can find in Ref. \cite{susyec} for more details
on this Lagrangian. In the following, only terms relevant to our
calculations are displayed.

\subsection{\label{gaugeboson}Gauge bosons}

The mass Lagrangian for the gauge bosons is given by
 \bea
\mathcal{L}_{mass}^{gauge}&=&(D^\mu\langle\rho\rangle)^+(D_\mu\langle\rho\rangle)+
(D^\mu\langle\chi\rangle)^+(D_\mu\langle\chi\rangle)\crn
&&+(\bar{D}^\mu\langle\rho'\rangle)^+(\bar{D}_\mu\langle\rho'\rangle)+
(\bar{D}^\mu\langle\chi'\rangle)^+(\bar{D}_\mu\langle\chi'\rangle),
\eea where \begin{eqnarray} D_\mu &=&\partial_\mu + ig T^a
V_{a\mu} +ig^\prime X T^9 B_\mu, \hs \bar{D}_\mu = \partial_\mu -
ig T^{a*} V_{a\mu} +ig^\prime X T^9 B_\mu.\end{eqnarray}

Let us define the charged gauge bosons as follows \be
W'^{\pm}_\mu\equiv\fr{1}{\sqrt{2}}(V_{1\mu}\mp iV_{2\mu}),\hs
Y'^\pm_\mu \equiv \fr{1}{\sqrt{2}}(V_{6\mu}\pm i V_{7\mu}).\ee The
mass matrix of the $W'_\mu$ and $Y'_\mu$ is obtained then
\begin{eqnarray}
M_{charged}^2  &=& \frac{g^2}{4}  \left(
                       \begin{array}{cc}
                         V^2+U^2& K   \\
                         K &  W^2+V^2   \\
                         \end{array}
                     \right),
\end{eqnarray}
where \bea V^2 &\equiv& v^2+v^{\prime 2},\hs W^2\equiv
w^2+w^{\prime 2},\hs U^2 \equiv u^2+u^{\prime 2}=t^2_\theta
W^2,\crn K &\equiv& uw+u^\prime w^\prime=t_\theta W^2,\hs t\equiv
g^\prime/g.\eea As in the previous work \cite{susyec}, we have
used \be t_{\theta} \equiv \fr{u}{w}=\fr{u'}{w'},
\label{ht2tan}\ee and $s_\theta \equiv \sin \theta$, $t_\theta
\equiv \tan \theta$, and so forth.

The physical gauge bosons are the SM-like $W^{\pm}$ and new gauge
boson $Y^{\pm}$: \be W_\mu = c_\theta W'_\mu-s_\theta Y'_\mu,\hs
Y_\mu=s_\theta W'_\mu +c_\theta Y'_\mu,\ee with the respective
masses: \bea m^2_{W}&=&\frac{g^2}{4}V^2, \hs
m^2_{Y}=\frac{g^2}{4}\left(V^2+U^2+W^2\right). \label{masswy}\eea
Therefore, the $\theta$ is the mixing angle of $W'-Y'$, which is
the same as in the case of non-supersymmetric model
\cite{haihiggs}. Because of the constraint (\ref{contraint}), the
mass of $W$ boson is identified with those of the SM, that is \be
 \sqrt{v^2+v'^2}\equiv v_{\mathrm{weak}}=246\ \mathrm{GeV}.\ee

For the remaining gauge vectors $(V_3,V_8, B,V_4,V_5)$, the mass
matrix in this basis is given by \bea
 M^2_{neutral} &=& \left(
 \begin{array}{cc}
       M_{mixing}^2&0\\
  0 & M^2_{V_5}
      \end{array}
   \right),
\eea where $V_5$ is decoupled with the mass \be
 M^2_{V_5}\equiv \frac{g^2}{4}\left(W^2+U^2\right),\label{klw5}\ee
while the mixing part $M^2_{mixing}$ of $(V_3,V_8,B,V_4)$ is equal
to \bea \frac{g^2}{4}\left(
   \begin{array}{cccc}
   U^2+V^2 & \frac{1}{\sqrt{3}}\left (U^2-V^2\right)
   & -\frac{2t}{3\sqrt{6}}\left(U^2+2V^2\right)
    & K \\
    & \frac{1}{3}\left(V^2+U^2+4W^2 \right) &
\frac{\sqrt{2} t}{9}\left(2V^2+2W^2-U^2 \right)& -\frac{1}{\sqrt{3}}K \\
&  & \frac{2t^2}{27}\left( 4V^2+U^2+W^2 \right) & -\frac{4t}{3\sqrt{6}}K \\
        &  &  &
         U^2+W^2 \\
                \end{array}
  \right) \label{M4neu}
\eea
 As in the non-supersymmetric version,
it can be checked that the  matrix (\ref{M4neu}) contains two {\it
exact} eigenvalues,
  the photon $A_\mu$ and new $V'_{4\mu}\sim V_{4\mu}$,
  such as
   \bea
   M^2_\gamma &=& 0, \crn
   M^2_{V^\prime_4}&=& \frac{g^2}{4}\left(U^2+W^2 \right).
  \label{xo} \eea
Due to the fact that $V^\prime_4$ and $V_5$ gain the same mass
[cf. (\ref{xo}) and (\ref{klw5})], it is worth noting that these
boson vectors have to be combined to produce the following
physical state \cite{haihiggs}\be X^0_\mu
\equiv\fr{1}{\sqrt{2}}(V^\prime_{4\mu}-iV_{5 \mu}),\label{dnx} \ee
with the mass \be m^2_X=\fr{g^2}{4}(U^2+W^2).\label{massx}\ee

Combining Eqs. (\ref{masswy}) and (\ref{massx}), as in the
non-symmetric version, we get
 the law of
Pythagoras \bea M^2_{Y}&=& M^2_{X}+M^2_{W}. \label{massrel} \eea
The eigenvectors of (\ref{M4neu}) are the same in
Ref.~\cite{haihiggs} of the non-supersymmetric version with unique
replacement of $u, v, w$ by $U, V, W$ (for details, see
\cite{susyec}). It is worth noting that because of the relation
(\ref{ht2tan}), the above diagonalization was eased. For
convenience in reading further the mixing matrix of the neutral
gauge bosons is given as follows \be \left(
  V_3, V_8,  B,  V_4
\right)^T= U\left(
  A,  Z,  Z',  V'_{4}
\right)^T, \ee where  \bea U=\left(%
\begin{array}{cccc}
  s_W & c_\va c_{\theta'}c_W  & s_\va c_{\theta'}c_W  & s_{\theta'}c_W \\
  -\fr{s_W}{\sqrt{3}} & \fr{c_\va(s^2_W-3c^2_Ws^2_{\theta'})
  -s_\va \la\kappa}{\sqrt{3}c_Wc_{\theta'}} &
  \fr{s_\va(s^2_W-3c^2_Ws^2_{\theta'})+
  c_\va \la \kappa}{\sqrt{3}c_Wc_{\theta'}} & \sqrt{3}s_{\theta'}c_W \\
  \fr{\kappa}{\sqrt{3}} & -\fr{t_W(c_\va\kappa
  +s_\va \la)}{\sqrt{3}c_{\theta'}} & -\fr{t_W(s_\va \kappa
 -c_\va \la)}{\sqrt{3}c_{\theta'}}  & 0 \\
  0 & -t_{\theta'}(c_\va\lambda
  -s_\va \kappa) & -t_{\theta'}(s_\va\la
  +c_\va \kappa) & \la
\end{array}%
\right)\label{gaugemx1},\\ s_{\theta'}\equiv
t_{2\theta}/(c_W\sqrt{1+4t^2_{2\theta}}),\hs
\kappa\equiv\sqrt{4c^2_W-1},\hs
\la\equiv\sqrt{1-4s^2_{\theta'}c^2_W}.\label{gaugemx}\eea

To finish this section, we mention again that the imaginary part
of the non-Hermitian bilepton $X^0$ is decoupled, while its real
part has the mixing among the neutral Hermitian gauge bosons such
as, the photon, the neutral $Z$ and the extra $Z'$.

\section{\label{sec:massspectrum} The Higgs sector revisited}
The supersymmetric Higgs potential takes the form \cite{susyec} \bea
V_{susyeco} &\equiv & V_{scalar} + V_{soft}\crn
 &=& \frac{\mu_{\chi}^2}{4}\left(\chi^\dagger\chi+\chi^{\prime\dagger}\chi^\prime\right)
+\frac{\mu_{\rho}^2}{4}\left(\rho^\dagger\rho+\rho^{\prime\dagger}\rho^\prime\right)
\crn &&+\frac{g^{\prime2}}{12} \left(-\frac{1}{3} \chi^{\dagger}
\chi + \frac{1}{3} \chi^{\prime \dagger} \chi^{\prime}+ \frac{2}{3}
\rho^{\dagger}\rho - \frac{2}{3} \rho^{\prime \dagger}\rho^{\prime}
\right)^2\crn & & +\frac{g^2}{8}(\chi^\dagger_i\lambda^b_{ij}\chi_j-
\chi^{\prime\dagger}_i\lambda^{*b}_{ij}\chi^\prime_j+
\rho^\dagger_i\lambda^b_{ij}\rho_j-
\rho^{\prime\dagger}_i\lambda^{*b}_{ij}\rho^\prime_j)^2\!\crn
&&+m^2_\rho\rho^\dagger\rho+m^2_\chi\chi^\dagger\chi
+m^2_{\rho^\prime}\rho^{\prime\dagger}\rho^\prime+
m^2_{\chi^\prime}\chi^{\prime\dagger}\chi^\prime. \label{p4} \eea
Assuming that the VEVs of neutral components $u,\ u',\ v,\ v',\ w$
and $w'$ are real, we expand the fields around the VEVs as
follows\bea \chi^T&=&\left(
           \begin{array}{ccc}
            \fr{u+S_1+iA_1}{\sqrt{2}}, & \chi^{-}, & \fr{w+S_2+iA_2}{\sqrt{2}} \\
           \end{array}
         \right), \hs  \rho^T = \left(
 \begin{array}{ccc}
  \rho_1^+, & \fr{v+S_5+iA_5}{\sqrt{2}}, & \rho_2^+ \crn
          \end{array}
      \right),\label{2}\\ {\chi^\prime}^T&=&\left(
           \begin{array}{ccc}
\fr{u^\prime+S_3+iA_3}{\sqrt{2}}, & \chi^{\prime
+}, & \fr{w^\prime+S_4+iA_4}{\sqrt{2}} \\
           \end{array}
         \right),\hs {\rho ^\prime}^T =\left(
                         \begin{array}{ccc}
                           \rho_1^{\prime -}, &
\fr{ v^\prime +S_6+iA_6}{\sqrt{2}}, & \rho_2^{\prime-} \\
\end{array}\right).\label{1}
\eea Requirement of vanishing the linear terms in fields, we get, at
the tree-level approximation, the following constraint equations
\bea \mu_\chi^2+ 4 m_\chi^2&=& -\frac{ g^{\prime 2}}{54}\left[
w^2-w^{\prime 2}+u^2-u^{\prime 2} + 2\left(v^{\prime 2}-v^2 \right)
\right] \nonumber
\\ && -\frac{g^2}{6}\left[2\left( u^2-u^{\prime 2}+w^2-w^{\prime
2}\right) +v^{\prime 2}-v^2 \right],\\ \mu_\rho^2+4 m_\rho^2 &=&
-\frac{2 g^{2 \prime}+9g^2}{54}\left[ 2\left(v^2-v^{2 \prime}
\right) +w^{2 \prime}-w^2 +u^{ \prime 2}-u^2\right],\eea \be
m_{\chi}^2+m^2_{\chi \prime} + \mu^2_{\chi} = 0,\ee \be
m_{\rho}^2+m^2_{\rho \prime} + \mu^2_{\rho} = 0,\ee \be \left(
w^2-u^2\right)u^\prime w^\prime = \left(w^{\prime 2} -u^{\prime
2}\right)uw. \label{tuyentinh}\ee It is noteworthy that Eq.
(\ref{tuyentinh}) implies the matching condition previously
mentioned in (\ref{ht2tan}). Consequently, the model contains a pair
of Higgs triplet $\chi$ and antitriplet $\chi^\prime$ with the VEVs
in top and bottom elements governed by the relation: $u/w
=u^\prime/w^\prime$.

The squared-mass matrix derived from (\ref{p4}) can be divided into
three $(6\times6)$ matrices respective to the charged, scalar and
pseudoscalar bosons. Note that there is no mixing among the scalar
and pseudoscalar bosons.

{\it Pseudoscalar sector}: \ben
\item
 There are two decoupled  massless particles: $A_5, A_6$.
\item Three  massless states are mixing of
\bea A_1^\prime & = & s_\bet A_1 -c_\bet  A_3, \\
 A_2^\prime & = & s_\bet A_2 - c_\bet A_4,\\
\varphi_A&=& s_\theta A_3^\prime +c_\theta A_4^\prime,
 \label{giavohuong1t} \eea
where \be
  A_3^\prime = c_\bet
A_1 + s_\bet A_3,\hs A_4^\prime = c_\bet A_2 + s_\bet  A_4,
\label{candDM}\ee with \be t_\bet \equiv
\frac{w}{w^\prime}.\label{theta1}\ee
 \item One massive eigenstate,
\be \phi_A= c_\theta A_3^\prime - s_\theta
A_4^\prime\label{candDM2},\ee with mass is equal to those of the $X$
bilepton \cite{susyec} \bea
m^2_{\phi_A}=\frac{g^2}{4}(1+t^2_\theta)(w^2+w^{\prime 2})= m^2_{X}.
\label{hx}\eea \een

Hence, in the pseudoscalar sector, we get five Goldstone bosons:
$A_5$, $ A_6$, $ A_1^\prime$,  $A_2^\prime$, $\va_A$ and one massive
$\phi_A$ having the mass equal to those of the bilepton $X$.

{\it Scalar sector}:

In this sector, six particles are mixing in terms of an $6 \times
6 $ squared-mass matrix. To study physical eigenvalues and
eigenstates, we change the basis to such
$\left(S_1^\prime,S_2^\prime,S_3^\prime,S_4^\prime,S_5^\prime,S_6^\prime
\right)$ as \bea \left(
 \begin{array}{c}
 S_1 \\
 S_2 \\
 S_3 \\
 S_4\\
 S_5\\
 S_6 \\
\end{array}
\right) &=&\left(
 \begin{array}{cccccc}
s_\theta & -c_\theta
& 0 & 0 & 0 & 0 \\
 c_\theta & s_\theta & 0 & 0 & 0 & 0 \\
0 & 0 & s_\theta  & -c_\theta  & 0 & 0 \\
 0 & 0 & c_\theta & s_\theta & 0 & 0 \\
 0& 0 & 0 & 0 & \frac{v^\prime}{\sqrt{v^2+v'^2}} & \frac{-v}{\sqrt{v^2+v^{\prime2}}} \\
0 & 0 & 0 & 0 & \frac{v}{\sqrt{v^2+v'^2}} & \frac{v'}{\sqrt{v^2+v^{\prime2}}}\\
\end{array}
\right)\left(
\begin{array}{c}
S_1^{\prime} \\
S_2^{\prime}\\
S_3^{\prime}\\
 S_4^{\prime} \\
 S_5^{\prime} \\
 S_6^{\prime} \\
\end{array}
 \right).
\eea For using further, we just introduce the following notation
\bea \left(
  \begin{array}{c}
    S_{1a}^\prime \\
    S_{3a}^\prime \\
    S_{6a}^\prime \\
  \end{array}
\right)= \left(
  \begin{array}{ccc}
    c_\beta &
s_\beta & 0 \\
    -s_\beta &
c_\beta & 0 \\
    0 & 0 & 1 \\
  \end{array}
\right)\left(
         \begin{array}{c}
           S_1^\prime\\
           S_3^\prime\\
           S_6^\prime \\
         \end{array}
       \right).
 \eea
With these combinations, we get the physical fields as follows \ben
\item Three massless fields: $S_5^\prime,  S_{1a}^\prime$, and
\bea \varphi_{S_{24}} & = & s_\beta S'_2 + c_\beta S'_4.\eea
\item Three massive fields corresponding to the masses:
\bea \phi_{S_{24}} & = & c_\beta S'_2 - s_\beta S'_4, \hs
m^2_{\phi_{S_{24}}} =\frac{g^2}{4}(1+t^2_\theta)(w^2+w'^2) =
m_{X}^2;\label{masshx}\\
\varphi_{S_{a36}} & = & s_\al S'_3 +c_\al S'_6,\\
&& \
 m^2_{\varphi_{S_{a36}}} = \frac{1}{2}
\left[m^2_{33a}+m^2_{66a}-
\sqrt{\left(m^2_{33a}-m^2_{66a}\right)^2 +4m^4_{36a}}
\right];\label{klvhnho}\\
\phi_{S_{a36}} & = &c_\al S'_3 -s_\al S'_6, \\
&& \
 m^2_{\phi_{S_{a36}}}  = \frac{1}{2}
\left[m^2_{33a}+m^2_{66a}+ \sqrt{\left(m^2_{33a}-m^2_{66a}\right)^2
+4m^4_{36a}} \right],\nn\eea where \bea m^2_{33a}&=&\fr{18 g^2 +
g'^2}{54c^2_\theta}(w^2 + w'^2),\hs m^2_{66a}=\fr{9g^2 +
2g'^2}{27}(v^2 + v'^2),\crn m^2_{36a}&=&\fr{(9g^2+2g'^2)\sqrt{(v^2 +
v'^2)(w^2 + w'^2)}}{54 c_\theta}\nn \eea and \be t_{2\al}\equiv
\fr{-2m^2_{36a}}{m^2_{66a}-m^2_{33a}}.\ee
 \een
From (\ref{klvhnho}), we get \be  m^2_{\varphi_{S_{a36}}} \simeq
\fr{h_1h_2-h_3^2} {h_1}(v^2+v'^2), \label{kln2} \ee where \bea
h_1\equiv\fr{18 g^2 + g'^2}{54c^2_\theta},\hs h_2\equiv\fr{9g^2 +
2g'^2}{27},\hs h_3\equiv\fr{9g^2 + 2g'^2}{54c^2_\theta}.\nn\eea
Taking into account $\alpha=\fr{e^2}{4\pi}=\fr{1}{128},\
s_W^2=0.2312,\ t=\fr{g'}{g}=\fr{3\sqrt{2}s_W}{\sqrt{4c^2_W-1}}$
\cite{dl} we have \be
 m_{\varphi_{S_{a36}}} \simeq
 91.4 \  \textrm{GeV}.
\ee This value is very closed to the lower limit of 89.8 GeV (95\%
CL) given in Ref. \cite{pdg} p. 32. It is interesting to note that
this mass is also closed to the $Z$ boson mass.

 Let us note that $\phi_A$ and
$\phi_{S_{24}}$ have the same mass, which can be combined to become
a physical neutral complex field
$H^0_X=(\phi_{S_{24}}+i\phi_A)/\sqrt{2}$ with mass equal to $m_X$ of
the neutral non-Hermitian gauge boson $X^0$.

 {\it Charged Higgs sector}:

In Ref.\cite{susyec}, to solve the characteristic equation for
charged Higgs sector, we had to use the approximation
(\ref{contraint}). In this sector we are revising the previous
work. Our result below is exact without any approximation.

In the base of ($\chi^{+}_a , \chi^{\prime
+}_a,\rho^+_{1a},\rho^+_{2a},\rho^{\prime +}_{1a}, \rho^{\prime
+}_{2a}$), the mass  matrix  becomes \cite{susyec} \bea
M^2_{a6charged} &=& \frac{g^2}{4}\left(
  \begin{array}{cccccc}
    m^2_{a11} & m^2_{a12} & 0 & m^2_{a14} & 0
    &m^2_{a16} \\
    & m^2_{a22} & 0 & 0 & 0 & 0 \\
     & & v^{\prime 2} & 0 & -vv^\prime & 0 \\
     &  &  &
    m^2_{a44} & 0 & -vv^\prime \\
     &  &  &  & v^2 & 0 \\
     &  &  &  & & m^2_{a66} \\
  \end{array}
\right), \label{LK}\eea where \bea m^2_{a11}&=& -c_{2\beta}
\left(\cot^2_\gamma-1 \right)v^{\prime 2},\hs
m^2_{a12}=-s_{2\beta} \left(\cot^2_\gamma-1 \right)v^{\prime
2},\nonumber \\  m^2_{a22}&=& c_{2\beta} \left(\cot^2_\gamma-1
\right)v^{\prime 2}+\left( 1+t^2_\beta
\right)\left( 1+t^2_\theta \right)w^{\prime 2}, \nonumber \\
m^2_{a14}&=& \sqrt{\left( 1+t^2_\beta \right)\left( 1+t^2_\theta
\right)}w^\prime v,\hs
 m^2_{a16}  = \sqrt{\left(
1+t^2_\beta \right)\left( 1+t^2_\theta \right)}w^\prime v^\prime,
\nonumber
\\ m^2_{a44}&=& \left(t^2_\beta -1 \right)\left(t^2_\theta+1
\right)w^{\prime 2}+v^{\prime 2}, \crn m^2_{a66}&=&-\left(t^2_\beta
-1 \right)\left(t^2_\theta+1 \right)w^{\prime 2}+v^{2},\eea with
\bea \cot_\gamma \equiv\frac{v}{v^\prime}.\eea Since the block
intersected by the third, fifth rows and columns is decoupled, it
can be diagonalized and this yields two eigenvalues as follows \bea
m^2_{\varrho_{1}^+}&=&\frac{g^2}{4}\left(
v^2+v^{\prime 2}\right) = m_W^2,\label{qhrow}\\
m^{2}_{\varrho^{+}_2}&=&0.\eea Here the Goldstone boson
$\varrho^+_2$ and Higgs boson $\varrho^+_1$ are, respectively,
defined by \bea \varrho_{1}^+&=& s_\ga \rho_{1a}^{+} -
c_\ga\rho_{1a}^{\prime +},\hs \varrho_2^+= c_\ga\rho_{1a}^+ +s_\ga
\rho_{1a}^{\prime +}.\label{nghiemtd}\eea Equation (\ref{qhrow})
shows that {\it one charged Higgs boson has the mass equal to
those of $W$ boson, i. e. $m_{\varrho^{\pm}_1} = m_{W^\pm}$}, this
result is in agreement with the experimental current limit $m
>79.3\ \mathrm{GeV},\ \mathrm{CL}=95\% $ \cite{pdg}.

The remaining part of ($\chi^{+}_a , \chi^{\prime
+}_a,\rho^+_{2a}, \rho^{\prime +}_{2a}$) is still mixing in terms
of an $4\times 4$ submatrix of (\ref{LK}). This matrix can be
diagonalized to get
eigenvalues as following \bea m^2_{ \zeta^+_1} &=& 0,\\
m^2_{\zeta^+_2} &=&
\frac{g^2}{4}\left[(t^2_\beta-1)(u^{\prime2}+w^{\prime2})-
(\cot^2_\gamma-1)v^{\prime2}\right], \label{ctm1}
\\ m^2_{\zeta^+_3}
&=&-m^2_{\vartheta^+_2}\label{ctm2},\\ m^2_{\zeta^+_4}&=&
\frac{g^2}{4}\left(U^2+V^2+W^2\right) = m^2_Y \label{ctm1t}\eea
and the corresponding eigenvalues\bea \zeta^+_1 &=&
\frac{\sqrt{(t_\beta^2+1)(u^{\prime2}+w^{\prime2})}}{\sqrt{\frac{4m^2_
{\vartheta^+_4}}{g^2}}}\chi_a^+ -\frac{v}{\sqrt{\frac{4m^2_
{\vartheta^+_4}}{g^2}}}\rho^+_{2a}-\frac{v^\prime}
{\sqrt{\frac{4m^2_ {\vartheta^+_4}}{g^2}}}\rho^{\prime+}_{2a},
\\ \zeta^+_2 &=&
\frac{1}{\sqrt{u^{\prime2}+w^{\prime2}+v^2}}\left[\frac{1}{\sqrt{1+t_\beta^2}}\left(v
\chi_a^+ + t_\beta
v\chi_a^{\prime+}\right)+\sqrt{u^{'2}+w^{'2}}\rho^+_{2a} \right],\\
 \zeta^+_3 &=&\frac{1}{\sqrt{v^{\prime2}+w^2+u^2}}\left[
 \frac{v^\prime}{\sqrt{1+t_\beta^2}} \left(t_\beta \chi^+_a
 -\chi^{\prime+}_a \right)+\sqrt{u^2+w^2}\rho_{2a}^{\prime +}
 \right],
\\  \zeta^+_4 &=&\frac{1}{\sqrt{k_1^2+k_2^2+k_2^2+1}}
\left(k_1\chi^+_a+k_2\chi^{\prime+}_a +k_3\rho^+_{2a}+
\rho^{\prime+}_{2a}\right),\eea where \bea
k_1&\equiv&-\frac{(t_\beta^2v^2-v^{\prime 2})\sqrt{u^{\prime
2}+w^{\prime
2}}}{v'\sqrt{1+t_\beta^2}(u^{\prime 2}+v^2+w^{\prime 2)}},\\
k_2&\equiv& \frac{\sqrt{u^2+w^2}\left[v^2+v^{\prime
2}+(1+t_\beta^2)(u^{2\prime}+w^{2\prime}) \right]}{v^\prime
\sqrt{1+t_\beta^2}(u^{\prime 2}+w^{2\prime}+v^2)},\\
k_3&\equiv&-\frac{v\left(v^{\prime 2}+u^2+w^2\right)}{v^\prime
\left( u^{\prime 2}+w^{\prime 2}+v^2 \right)}.\eea As in the gauge
boson sector, from (\ref{masshx}), (\ref{qhrow}) and (\ref{ctm1t}),
we get again the law of Pythagoras \bea
 m^2_{\zeta^+_4} &=& m^2_{H_X^0}+m^2_{\varrho_{1}^+}. \label{masshiggs} \eea

It is easy to check that the physical field $\zeta^+_1$ is Goldstone
bosons and charged Higgs boson $\zeta_4$ has the mass equal to those
of $Y$. This matrix also gives us two physical fields $\zeta^+_2$
and $\zeta^+_3$ with their mass are the same value but opposite
sign. Therefore, one of them can be identified with tachyon fields.

From (\ref{ctm1}) and  (\ref{ctm2}), to cancel the tachyon field, we
have to put the condition \be
(t^2_\beta-1)(u^{\prime2}+w^{\prime2})-(\cot^2_\gamma-1)v^{\prime2}=0.
\label{tachcond}\ee This yields  \be 1 + \fr{u^{\prime 2}}{w^{\prime
2}} = \fr{v^2 - v^{\prime 2}}{w^2 - w^{\prime
2}}.\label{tachcond2}\ee

This means that in the limit $w^\prime  \gg u^\prime $, we have the
splitting formula
 \be
w^2 - w^{\prime 2} = v^2 - v^{\prime 2} \leq 246^2  \
\textrm{GeV}^2. \label{tachcond3}\ee It is noteworthy that the
relation (\ref{tachcond3}) is very good addition to (\ref{massrel}).

Finally, let us summarize the physical fields of the scalar sector
in the model. There are eight neutral massless particles: five
pseudoscalars $A_5$, $A_6$, $A'_1$, $A'_2$, $\varphi_A$, and three
scalars $S'_5$, $\varphi_{S_{24}}$, $S'_{1a}$. There is one complex
neutral Higgs $H^0_X$ with mass equal to those of the bilepton
$m_X$, and two massive scalars $\varphi_{S_{a36}}$,
$\phi_{S_{a36}}$. There are four charged massless scalar fields
$\varrho^{\pm}_2,\ \zeta^\pm_1,\ \zeta^\pm_2$ and $\varrho^{\pm}_3$,
and two massive charged bosons $\varrho^{\pm}_1$ and $\zeta^\pm_4$
with masses equal to that of the $W$ boson and the bilepton $Y$,
respectively: $m_{\varrho^{+}_1} = m_W,\  m_{\zeta^+_4} = m_Y$.

\section{Higgs-gauge boson couplings}
\label{susyhg}

With above content of Higgs sector, we can now calculate the
Higgs-gauge boson interactions. These interactions exist in part
from
 \begin{eqnarray}
\mathcal{L}_{kinetic} &=& (D^\mu \chi)^+ D_\mu \chi +(D^\mu
\rho)^+ D_\mu
 \rho+ (\bar{D}^\mu \chi^{\prime})^+ \bar{D}_\mu \chi^{\prime} +
 (\bar{D}^\mu \rho^{\prime})^+
 \bar{D}_\mu \rho^{\prime}.\label{Lnosusy1} \end{eqnarray}
In this paper, the gauge bosons are limited to be the gauge bosons
of Glashow-Weinberg-Salam model, i.e., photon, $W$ and $Z$ bosons.
Using mixing matrices given in Appendix \ref{pl1}, we are able to
get interactions of the physical fields.

Despite mixing, electromagnetic interactions are unchanged \be i e
A^\mu H^- \stackrel{\leftrightarrow}{\pa_\mu}H^+, \hs
H^-=\varrho^-_1,\varrho^-_2,\zeta^-_1,\zeta^-_2,\zeta^-_3,\zeta^-_4.
\label{ttdt}\ee

For the $W$ boson, we get  couplings of pair $W^+W^-$ with  neutral
Higgs bosons presented in Table \ref{tab3}.

The interactions of single $W$ boson with two Higgs bosons are
presented in Table \ref{tab4}, where $m_{x,y}, x,y = 1,2,3,4$ are
given in Appendix \ref{pl1}. Other vertices are \bea \mathcal{V}(W^-
\varrho^+_1 A_5)&=&- \mathcal{V}(W ^-\varrho^+_2
A_6)=\fr{gv'}{2\sqrt{v^2+v'^2}}, \crn \mathcal{V}(W ^-\varrho^+_2
A_5)&=&
\mathcal{V}(W^- \varrho^+_1 A_6)=\fr{gv}{2\sqrt{v^2+v'^2}}, \\
\mathcal{V}(W^- \varrho^+_2 S'_5)&=& \fr{-1}{c_\alpha}
\mathcal{V}(W^- \varrho^+_1
\varphi_{S_{a36}})=\fr{1}{s_\alpha}\mathcal{V}(W^-
\varrho^+_1 \phi_{S_{a36}})=-\fr{ig vv'}{v^2+v'^2},\\
\mathcal{V}(W ^-\varrho^+_1 S'_5)&=& \fr{1}{c_\alpha}
\mathcal{V}(W^- \varrho^+_2
\varphi_{S_{a36}})=\fr{-1}{s_\alpha}\mathcal{V}(W ^-\varrho^+_2
\phi_{S_{a36}})=\fr{ig (v^2-v'^2)}{v^2+v'^2}. \eea

Non-zero quartic couplings of  pair $W^+W^-$ with two Higgs bosons
are given in Table \ref{tab5}. Addition to this table, we have also
five interactions \bea && \mathcal{V}(W^+ W^-  A_5 A_5)
=\mathcal{V}( W^+ W^- A_6 A_6) = \mathcal{V}(W^+W^- S'_5 S'_5)\crn
&&=\mathcal{V}( W^+W^- H^0_X  H^{0*}_X)=
\mathcal{V}(W^+W^-\varrho^+_1\varrho^-_1) =
\mathcal{V}(W^+W^-\varrho^+_2\varrho^-_2).\eea

For the neutral $Z$ boson, the triple coupling of single $Z$ with
two charged Higgs bosons are presented in Table \ref{tab6}, where
$U_{22}$, ... are elements in the mixing matrix of the neutral
gauge bosons given in (\ref{gaugemx1}). We have also
 \bea
\mathcal{V}(Z  \varrho^-_1  \varrho^+_1) = \mathcal{V}(Z \varrho^-_2
\varrho^+_2).
 \eea
The notations are given by
 \bea f_1 &=&
u(3U_{12}+\sqrt{3}(U_{22}-2s_W\sqrt{\fr{1}{3-4s^2_W}} U_{32}))+ 3w
U_{42},\\
f_2&=&3U_{12}-\sqrt{3}(U_{22}+4s_W\sqrt{\fr{1}{3-4s^2_W}}U_{32}),\\
 f_3&=&-3U^2_{12}-2\sqrt{3}U_{12}(U_{22}-2s_W\sqrt{\fr{1}{3-4s^2_W}}
U_{32})\crn
&&+3U_{22}(U_{22}+4s_W\sqrt{\fr{1}{3-4s^2_W}}U_{32}),\\
f_4 &=&3U^2_{12}+U^2_{22}+
\fr{8s_WU_{22}U_{32}}{\sqrt{3-4s^2_W}}-\fr{16s^2_WU^2_{32}}{-3+4s^2_W}\crn
&& -2\sqrt{3}U_{12}(U_{22}+\fr{4s^2_W}{\sqrt{3-4s^2_W}}U_{42}),\\
f_5&=&3U^2_{12}+U^2_{22}-\fr{4s_WU_{22}U_{32}}{\sqrt{3-4s^2_W}}
-\fr{4s^2_WU^2_{32}}{-3+4s^2_W}\crn
 &&-2\sqrt{3}U_{12}(-U_{22}+\fr{2s_WU_{32}}{\sqrt{3-4s^2_W}})+3U^2_{42},\\
f_6&=&4U^2_{22}+\fr{8s_WU_{22}U_{32}}{\sqrt{3-4s^2_W}}
+3U^2_{42}-\fr{4s^2_W}{-3+4s^2_W}U^2_{32},\\
f_7 &=&3uU_{42}-2\sqrt{3}w(U_{22}+s_W\sqrt{\fr{1}{3-4s^2_W}}
U_{32}),\\
f_8  &=&  u^2[3U_{12}+\sqrt{3}(U_{22}-
 2s_W\sqrt{\fr{1}{3-4s^2_W}}U_{32})]\crn
 &&+ 6uw U_{42}-
2\sqrt{3}w^2(U_{22}+s_W\sqrt{\fr{1}{3-4s^2_W}} U_{32}).
 \eea
Similarly, trilinear coupling of the single $Z$ with two neutral
Higgs bosons are given in Table \ref{tab7}.

The triple couplings of pair $ZZ$ with one scalar Higgs boson are
given in Table \ref{tab8a}, where \bea
a_1&=&(a_{11}^{2}+U_{42}^{2})u+(a_{11}+a_{33})U_{42}w, \\
a_2&=&(a_{33}^{2}+U_{42}^{2})w +(a_{11}+a_{33})U_{42}u, \\
a_3&=&(a_{11}^{2}+U_{42}^2)u'-(a_{11}+a_{33})U_{42}w',\\
a_4&=&(a_{33}^{2}+U_{42}^2)w' -(a_{11}+a_{33})U_{42}u', \\
a_{11}&=&U_{12}+\frac{1}{\sqrt{3}}U_{22}-\frac{t}{\sqrt{3}}\sqrt{\frac{2}{3}}U_{32},\\
a_{33}&=&-\frac{2}{\sqrt{3}}U_{22}-\frac{t}{\sqrt{3}}\sqrt{\frac{2}{3}}U_{32}.\\
\eea

Non-zero quartic couplings of pair $ZZ$ with two scalar Higgs bosons
are  given in Table \ref{tab9}.

Non-zero quartic of pair $ZZ$ with two charged Higgs bosons are
presented in Table \ref{tab11}.

Another interaction is \bea \mathcal{V}(Z Z \rho_1  \rho_1)=
\mathcal{V}(Z Z \rho_2 \rho_2) \eea

In the special limit \bea && u = u',\hs w=w',\hs v'=0, \crn && w,
w'\gg  u, u', v, v', \label{smlimit}\eea the effective couplings are
summarized in Table \ref{tabeff}.
\begin{table}
\bc \caption{\label{tabeff}The non-zero coupling constants in the
effective limit.}

\vs

\begin{tabular}{|c|c|c|c|} \hline
Vertex  & Coupling  & Vertex & Coupling \\ \hline $W^+W^-S'_5$ &
$g^2vs_\gamma$  & $ZW^+\zeta^-_4$&$\fr{g^2u}{2\sqrt{2}c_W}$
\\
$W^+W^-\varphi_{S_{a36}}$ & $-\fr{g^2vc_{2\gamma}}{2c_\gamma}$&
$ZA_5\varphi_{S_{a36}}$&$-\fr{gc_\gamma}{2c_W}$
\\
$W^-\varrho^+_1\varphi_{S_{a36}}$ & $\fr{igs_{2\gamma}}{2}$  &
$ZA_6\varphi_{S_{a36}}$&$-\fr{gs_\gamma}{2c_W}$
\\ $W^-\varrho^+_1S'_5$ & $\fr{igc_{2\gamma}}{2}$&
$ZA_5S'_5$&$\fr{gs_\gamma}{2c_W}$
\\ $W^-\varrho^+_1 A_5$ & $\fr{gs_\gamma}{2}$  &
$ZA_6S'_5$&$-\fr{gc_\gamma}{2c_W}$
\\ $W^-\varrho^+_1 A_6$ & $\fr{gc_\gamma}{2}$ &
$ZA'_1\varphi_{S_{24}}$&$\fr{g}{2c_W}$
\\ $W^-\varrho^+_2\varphi_{S_{a36}}$ &  $\fr{igc_{2\gamma}}{2}$&
$ZZ\varphi_{S_{24}}$&$-\fr{g^2u}{2\sqrt{2}c^2_W}$
\\ $W^-\varrho^+_2S'_5$ & $-\fr{igs_{2\gamma}}{2}$&
$ZZ\varphi_{S_{a36}}$&$-\fr{g^2vc_{2\gamma}}{2c^2_Wc_\gamma}$
\\ $W^-\varrho^+_2A_6$ & $-\fr{gs_\gamma}{2}$ &
$ZZS'_5$&$\fr{g^2vs_\gamma}{c^2_W}$
\\ $W^-\varrho^+_2A_5$ & $\fr{gc_\gamma}{2}$ & $Z\Psi^+\Psi^-$&
$-\fr{igs^2_W}{2c_W},\ \Psi=\zeta_2, \zeta_3$
\\ $W^-\zeta^+_1 H^0_X$ & $\fr{ig}{\sqrt{2}}$ & $Z\Psi^+\Psi^-$ &
$\fr{igc_{2W}}{2c_W},\ \Psi=\varrho_1, \varrho_2, \zeta_1, \zeta_4 $
\\ $W^-\zeta^+_4\varphi_{S_{24}}$ & $\fr{ig}{2}$&
$ZZ\Psi^-\Psi^+$&$\fr{2g^2s^4_W}{c^2_W},\ \Psi=\zeta_2, \zeta_3$
\\ $W^-\zeta^+_4 A'_1$ & $-\fr{g}{2}$ &
$ZZ\Psi^-\Psi^+$&$\fr{g^2c^2_{2W}}{2c^2_W},\ \Psi=\varrho_1,
\varrho_2, \zeta_1, \zeta_4$
\\ $AW^+\varrho_2^-$ &$ -\fr{e^2v}{2s_Wc_\gamma}$& $WWHH$ &
$\fr{g^2}{2},\ H=A'_1, \varphi_{S_{a36}}, S'_5, A_6, H^0_X, \zeta_1,
\zeta_4, \varrho_1, \varrho_2 $
\\ $ZW^+\varrho_2^-$& $\fr{e^2v}{2c_Wc_\gamma}$&
$ZZHH$&$\fr{g^2}{2c^2_W},\ H=S'_5, \varphi_{S_{24}}, H^0_X,
\varphi_{S_{a36}}, \phi_{S_{a36}}, A'_1, A_5, A_6$
\\ \hline
\end{tabular}
\ec
\end{table} From (\ref{nghiemtd}) and Appendix
\ref{pl3} we get the following limit for physical fields \bea &&
S_5 \rightarrow - \varphi_{S_{a36}},\hs \rho^+_1
 \rightarrow - \varrho^+_2. \label{smlimit2} \eea
Therefore, the Higgs triplet responsible for the second step of
symmetry breaking $\rho$ can be represented as \be
 \rho \Rightarrow \left(
 \begin{array}{c}
 -\varrho^+_2 \\
\fr{v-\varphi_{S_{a36}}+iA_5}{\sqrt{2}}\\
 \zeta^+_2\\
 \end{array}
 \right)\label{cautruc1}\ee
Remind that both $\varrho^+_2$ and $A_5$ are massless. By the
Table \ref{tabeff}, we can identify them as Goldstone bosons for
the $W$ and $Z$ ones (neglecting the minus sign), respectively.
This yields
 \be
 \rho \Rightarrow \left(
 \begin{array}{c}
 G_{W^+} \\
\fr{v +  h + i G_Z}{\sqrt{2}}\\
\zeta^+_2\\
\end{array}
\right)\label{cautruc2}\ee Hence, all the effective couplings of the
gauge bosons with scalar fields of the SM can be recovered, which
most of them are presented in Table \ref{tab14}.
\begin{table} \bc \caption{\label{tab14}The SM coupling constants in
the effective limit.}

\vs

\begin{tabular}{|c|c|c|c|}\hline
Vertex  & Coupling  & Vertex & Coupling \\ \hline $WWhh$ &
$\fr{g^2}{2}$  & $G_WG_WA$ & $ie$
\\
$WWh$ & $\fr{g^2}{2}v$  & $WWG_ZG_Z$ & $\fr{g^2}{2}$
\\
$WG_W h$ & $-\fr{ig}{2}$  &$WWG_WG_W$ & $\fr{g^2}{2}$
\\
$WG_W G_Z$ & $\fr{g}{2}$  & $ZZh$ & $\fr{g^2}{2c^2_W}v$\\
$ZZhh$ & $\fr{g^2}{2c^2_W}$ & $ZZG_Z G_Z$ & $\fr{g^2}{2c^2_W}$ \\
$AWG_W$ & $\fr{g^2}{2}v s_W$ & $Z W G_W$ & $-\fr{g^2}{2}vs_W t_W$
\\
$ZG_Z h$ & $-\fr{g}{2c_W}$ & $ZG_W G_W$ & $\fr{ig}{2c_W}(1-2s^2_W)$
\\ \hline
\end{tabular}\ec
\end{table}

In principle, we cannot put $v^\prime =0$. The above analysis just
shows that our calculations are correct.

\section{
 Production of charged $\zeta_4^\pm$ via $WZ$ fusion at  LHC}
\label{chargeds}

The possibility to detect the neutral Higgs boson in the minimal
version at $e^+ e^-$ colliders was considered in~\cite{mal} and
production of the SM-like neutral Higgs boson in the 3-3-1 model
with right-handed neutrinos at the CERN LHC was considered in
Ref.\cite{ninhlong}. The decay and production at the CERN LHC of the
bilepton charged Higgs in the non-supersymmetric version of the
considering was given in Ref. \cite{dls1}. This section is devoted
to the decay modes and production of the charged $\zeta_4^\pm$ at
the CERN LHC.

Let us first discuss on the mass of this Higgs boson. Equation
(\ref{ctm1t}) gives us a connection between its mass and those of
the singly-charged bilepton $Y$. The bilepton mass limit can be
obtained from the ``wrong"  muon decay $\mu^- \rightarrow e^-
\nu_e \tilde{\nu}_\mu$ mediated, at the tree level, by both the
$W$ and the $Y$ boson. Taking into account of the famous
experimental data \cite{pdg}
 \be R_{muon} \equiv
\fr{\Ga(\mu^- \rightarrow e^- \nu_e \tilde{\nu}_\mu)}{\Ga(\mu^-
\rightarrow e^- \tilde{\nu}_e \nu_\mu)}
 < 1.2 \% \hs  \textrm{90 \% \ CL} \label{wrdecayrat}
 \ee we get the constraint: $ R_{muon} \simeq \fr{M_W^4}{M_Y^4}$.
 Therefore, it follows that $M_Y \geq 230 $ GeV. This bound is
 consistent with that followed from the oblique consideration in
 Ref. \cite{longinami}.
However, the stronger  bilepton mass bound has been derived from
consideration of experimental limit on  lepton-number violating
charged lepton decays \cite{tullyjoshi} of 440 GeV.

Taking into account that, in the effective approximation,
$\zeta^-_4$ is the bilepton, we get the dominant decay channels as
follows \bea \zeta^-_4 & \rightarrow & \left\{%
\begin{array}{l}
  l  \nu_l, \hs U^c d,\hs u^c D, \\
 Z W^-,\hs \tilde{H}^0 \tilde{W}^-.
\end{array}%
\right.\label{modes} \eea

Assuming that masses of the exotic quarks $(U, D_\al)$ and both
gaugino and Higgsino  are larger than $M_{\zeta^\pm_4}$, we come to
the fact that the hadron and sparticle modes are absent in the decay
of the charged Higgs boson. Because the Yukawa couplings of
$\zeta_4^\pm l^\mp \nu$ are very small, the coupling of a
singly-charged Higgs boson ($\zeta^\pm_4$) with the weak gauge
bosons, $\zeta^\pm_4 W^\mp Z$, can dominate. Note that the charged
Higgs bosons in doublet models such as the two-Higgs doublet model
or the minimal supersymmetric standard model, have both hadronic and
leptonic modes~\cite{roy}. This is a specific feature of the model
under consideration. It is of particular importance for the
electroweak symmetry breaking. Its magnitude is directly related to
the structure of the extended Higgs sector under global
symmetries~\cite{glob}. This coupling can appear at the tree level
in models with scalar triplets, while it is induced at the loop
level in multi scalar doublet models. The coupling, in our model,
differs from zero at the tree level due to the fact that the
$\zeta^\pm_4$ belongs to a triplet.

Thus, for the charged Higgs boson $\zeta^\pm_4$, it is important
to study the couplings given by the interaction Lagrangian \be
\mathcal{L}_{int} =  f_{ZW\zeta_4}\zeta_4^\pm W_\mu^\mp Z^\mu,
\label{potenn50} \ee where $f_{ZW\zeta_4}$, at tree level,  is
given in Table \ref{tab13}. The same as in~\cite{kame}, the
dominant rate is due to the diagram connected with the $W$ and $Z$
bosons. Putting necessary matrix elements in Table \ref{tab13} ,
we get \bea
f_{ZW\zeta_4}=\fr{g^2w^2t_\theta}{2(1+t_\beta^2)}\fr{v^2(2t^2_\gamma-t^2_\beta+1)
+\fr{1+t^2_\beta}{t^2_\beta}(u^2+w^2)}{X}\\
\times\fr{s_\varphi\sqrt{(4c^2_W-1)(1+4t^2_{2\theta})}-c_\varphi}
{\sqrt{c^2_W+t^2_{2\theta}(4c^2_W-1)}\sqrt{1+4t^2_{2\theta}}}
 \eea
 where
\bea X^2&&=v^4t^2_\gamma
V^2+\fr{u^2+w^2}{t^2_\beta}v^4(t^2_\beta+t^4_\gamma+2t^2_\gamma
t^2_\beta+2t^2_\gamma)+\\
&&\fr{(u^2+w^2)^2}{t^4_\beta}v^2(t^4_\beta +t^2_\gamma+2t^2_\gamma
t^2_\beta+2t^2_\beta)+ \fr{(u^2+w^2)^3}{t^6_\beta}(1+t^2_\beta)
\eea

Thus, the form factor, at the tree-level, is obtained by \bea
F\equiv\fr{f_{ZW\zeta_4}}{gM_W}=\fr{w^2t_\theta}{V(1+
t_\beta^2)}\fr{v^2(2t^2_\gamma-t^2_\beta+1)
+\fr{1+t^2_\beta}{t^2_\beta}(u^2+w^2)}{X}\\
\times\fr{s_\varphi\sqrt{(4c^2_W-1)(1+4t^2_{2\theta})}-c_\varphi}
{\sqrt{c^2_W+t^2_{2\theta}(4c^2_W-1)}\sqrt{1+4t^2_{2\theta}}}\label{fe}\eea
The decay width of $\zeta_4^\pm\rightarrow W^\pm_iZ_i$, where
$i=L,\ T$ represent respectively the longitudinal and transverse
polarizations, is given by~\cite{kame}
 \bea \Gamma(\zeta_4^\pm\to W^\pm_i Z_i) = M_{\zeta_4^\pm}
 \frac{\la^{1/2}(1, w, z)}{16\pi} |M_{ii}|^2,\eea where
$\la(1,w,z)=(1-w-z)^2-4wz$, $w=M^2_W/M^2_{\zeta^\pm_4}$ and
$z=M^2_Z/M^2_{\zeta^\pm_4}$. The longitudinal and transverse
contributions are given in terms of $F$ by \bea |M_{LL}|^2 &=&
\frac{g^2}{4 {z}}
      (1-{w}-{z})^2 \left|F
             \right|^2, \\
|M_{TT}|^2 &=& 2 g^2
     {w} |F|^2. \eea For the case of $M_{\zeta_4^\pm} \gg M_{Z}$, we have
$|M_{TT}|^2/|M_{LL}|^2 \sim 8 M_W^2 M_Z^2/M_{\zeta_4^\pm}^4$ which
implies that the decay into a longitudinally polarized weak boson
pair dominates that into a transversely polarized one.

Next, let us study the impact of the $\zeta^\pm_4 W^\mp Z$ vertex on
the production cross section of $pp \rightarrow W^{\pm*} Z^* X
\rightarrow \zeta^\pm_4 X$ which is a pure electroweak process with
high $p_T$ jets going into the forward and backward directions from
the decay of the produced scalar boson without color flow in the
central region. The hadronic cross section for $pp \to \zeta_4^\pm
X$ via $W^\pm Z$ fusion is expressed in the effective vector boson
approximation~\cite{kane} by 
\bea \sigma_{\rm eff}(s,M_{\zeta_4^\pm}^2) \simeq \frac{16\pi^2 }{
\lambda(1,w,z) M_{\zeta_4^\pm}^3} \sum_{\lambda=T,L}
\Gamma(\zeta_4^\pm \to W^\pm_\lambda Z_\lambda)
       \tau \left.\frac{d \mathcal{ L}}{d \tau}
      \right|_{pp/W^\pm_\lambda Z_\lambda},
\eea where $\tau=M_{\zeta_4^\pm}^{2}/s$, and \bea
 \left.\frac{d \mathcal{
L}}{d \tau}\right|_{pp/W^\pm_\lambda Z_\lambda} = \sum_{ij}
\int_\tau^{1} \frac{d\tau'}{\tau'} \int_{\tau'}^{1} \frac{d x}{x}
f_i(x) f_j(\tau'/x) \left.\frac{d \mathcal{ L}}{d
\xi}\right|_{q_iq_j/W^\pm_\lambda Z_\lambda}, \eea
 with
$\tau'=\hat{s}/s$ and $\xi=\tau/\tau'$. Here $f_i(x)$ is the
parton structure function for the $i$-th quark, and \bea \left.
\frac{d \mathcal{ L}}{d\xi} \right|_{q_iq_j/W_T^\pm Z_T}
&=&\frac{c}{64\pi^4} \frac{1}{\xi} \ln \left(
\frac{\hat{s}}{M_W^2} \right) \ln \left( \frac{\hat{s}}{M_Z^2}
\right) \crn && \times \left[ (2+\xi)^2 \ln
(1/\xi)-2(1-\xi)(3+\xi)
\right],\\
\left. \frac{d  \mathcal{ L}}{d\xi} \right|_{q_iq_j/W_L^\pm Z_L}
&=&\frac{c}{16\pi^4} \frac{1}{\xi} \left[ (1+\xi) \ln
(1/\xi)+2(\xi-1) \right], \eea
 where
$c=\fr{g^4c^2_\theta}{16c^2_W}\left[g_{1V}^{2}(q_j)+g_{1A}^{2}(q_j)\right]$
with $g_{1V}(q_j)$, $g_{1A}(q_j)$ for quark $q_j$ are given in
Table I of Ref.~\cite{haihiggs}.

\begin{figure}
\begin{center}
\includegraphics[width=10cm,height=7cm]{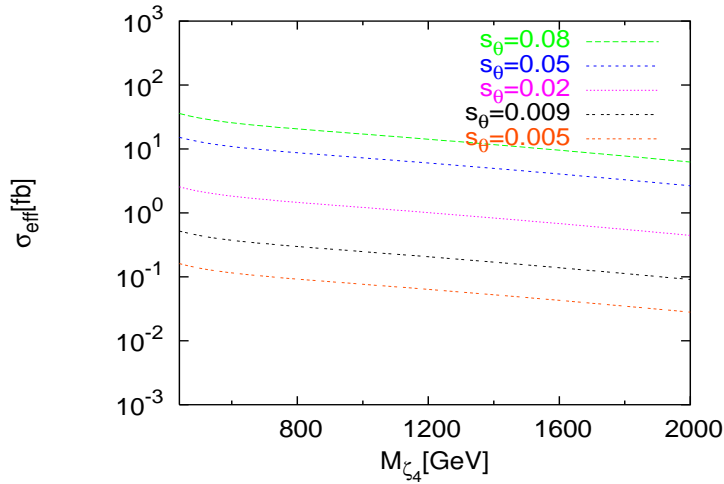}
\caption{\label{plseff}Hadronic cross section for production of
charged $\zeta_4^\pm$ via $WZ$ fusion as a function of the charged
Higgs boson mass for five cases of $\sin\theta$.}
\end{center}
\end{figure}

Using CTEQ6L~\cite{cteq6}, in Fig. \ref{plseff}, we plot
$\sigma_{\rm eff}(s,M_{\zeta_4^\pm}^2)$ at $\sqrt{s} = 14 \
\mathrm{TeV}$ as a function of $M_{\zeta_4}$ in range of 440---2000
GeV, where the parameters in the $F$ factor are set as follows
$v'=0$, $v=246$ GeV, $t_\beta=1$, and $t_\va$ obtained from Ref.
\cite{haihiggs}.
 If the mass of the charged Higgs boson is in range of
440 GeV and $s_\theta = 0.08$, the cross section can exceed 35.8
$fb$: i.e., 10740 of $\zeta_4^\pm$ can be produced at the integrated
LHC luminosity of $300\ fb^{-1}$.

\section{\label{concl} Conclusions}

In this paper we have explored the Higgs sector of the
supersymmetric  economical 3-3-1 model and found more new
interesting features in this sector. We have revised the charged
Higgs sector, i.e., the exact eigenvalues and states of the
charged Higgs fields were obtained without any approximation. In
this model, there are three Higgs bosons having  masses equal to
that of the gauge bosons and one neutral complex Higgs boson with
mass of the neutral non-Hermitian bilepton $X_0$. Therefore,  as
in the gauge sector, we get the law of Pythagoras among Higgs
boson masses: $
 m^2_{\zeta^+_4} = m^2_{H_X^0}+m^2_{\varrho_{1}^+}$.

There is one scalar  boson with mass of 91.4 GeV, which is closed to
the $Z$  boson mass and in good agreement with present limit: 89.8
GeV at 95\% CL.

The mass matrix of charged Higgs bosons gives two physical fields
$\zeta^+_2$ and $\zeta^+_3$ with their square  mass  are the same
value but opposite sign. To solve this problem, we have got very
interesting relation which leads to $w  \simeq w^\prime, u \simeq
u^\prime $ in high mass limit.

In the model under consideration,  at the tree level, the lightest
Higgs boson is the charged with the mass equal to those of the $W$
boson. This is in agreement with the current experimental limit:
79.3 GeV at 95\% CL.

It is worth mentioning that the Higgs sector in this model is very
constrained. At the tree level, we cannot fix {\it only} one heavy
scalar Higgs boson $\phi_{S_{a36}}$ with mass, while all remaining
fields gain masses of the gauge bosons in the model. This is nice
feature of the supersymmetric version.

The interactions among the standard model gauge bosons and scalar
fields  in the framework of the supersymmetric economical 3-3-1
model are also presented. From these couplings, all scalar fields
including the neutral scalar $h$ and the Goldstone bosons can be
identified and their couplings with the usual gauge bosons such as
the photon, the charged $W^\pm$ and the neutral $Z$, without any
additional condition, are recovered.

Despite the mixing among the photon with the non-Hermitian neutral
bilepton $X^0$ as well as with the $Z$ and the $Z'$ gauge bosons,
the electromagnetic couplings remain unchanged.

After all we focused attention to the singly-charged Higgs boson
$\zeta^\pm_4$ with mass equal to the bilepton mass $M_Y$. Mass of
the $\zeta^\pm_4$  is estimated to be larger than 440 GeV. This
boson, in difference with those arisen in the Higgs doublet models,
does not have the hadronic and leptonic decay modes. The trilinear
coupling $ZW^\pm \zeta_4^\mp$ which differs, at the tree level,
while the similar coupling of the photon $\ga W^\pm \zeta^\mp_4$ as
expected, vanishes. If the mass of the above mentioned Higgs boson
is in range of 440 GeV, however, the cross section can exceed
35.8~$fb$: i.e., 10740 of $\zeta_4^\pm$ can be produced at the CERN
LHC for the luminosity of 300~$fb^{-1}$.
 By measuring this process we can obtain useful information to
determine the structure of the Higgs sector.

LEPII placed the problem of Higgs physics at the forefront of
supersymmetry phenomenology.  While earlier one might have viewed
the Higgs fields as just one of many features of low energy
supersymmetric models, the constraints on the Higgs mass are now
problematic. In the model under consideration, the Higgs bosons
gain masses equal to that of the gauge bosons. This feature
deserves further studies.

\section*{Acknowledgments}
 The work was also supported in part by National Council
for Natural Sciences of Vietnam under grant  No: 410604.
\\[0.3cm]

\appendix

\section{Mixing matrix for neutral scalars}
\label{pl1}

\bea \left(
 \begin{array}{c}
 S_1 \\
 S_2 \\
 S_3 \\
 S_4\\
 S_5\\
 S_6 \\
\end{array}
\right) &=&\left(
 \begin{array}{cccccc}
c_\beta s_\theta & -s_\beta c_\theta
& -c_\beta c_\theta & -s_\alpha s_\beta s_\theta &  -c_\alpha s_\beta s_\theta & 0 \\
 c_\beta c_\theta & s_\beta s_\theta & c_\beta s_\theta & -s_\alpha s_\beta c_\theta
 & -c_\alpha s_\beta c_\theta & 0 \\
s_\beta s_\theta & -c_\beta c_\theta & s_\beta c_\theta  &
s_\alpha c_\beta s_\theta
& c_\alpha c_\beta s_\theta & 0 \\
 s_\beta c_\theta & c_\beta s_\theta & -s_\beta s_\theta & s_\alpha c_\beta c_\theta
 & c_\alpha c_\beta c_\theta & 0 \\
 0& 0 & 0 & -c_\alpha c_\gamma & s_\alpha c_\gamma & s_\gamma \\
0 & 0 & 0 & c_\alpha s_\gamma & -s_\alpha s_\gamma & c_\gamma\\
\end{array}
\right)\left(
\begin{array}{c}
S'_{1a} \\
\varphi_{S_{24}}\\
\phi_{S_{24}}\\
\varphi_{S_{a36}} \\
 \phi_{S_{a36}} \\
 S'_5 \\
\end{array}
 \right)
\eea \section{Mixing matrix for neutral pseudoscalars} \label{pl2}
\bea \left(
 \begin{array}{c}
 A_1 \\
 A_2 \\
 A_3 \\
 A_4\\
\end{array}
\right) &=&\left(
 \begin{array}{cccc}
s_\beta  & 0 & c_\beta s_\theta & c_\beta c_\theta  \\
0 & s_\beta & c_\beta c_\theta & -c_\beta s_\theta  \\
-c_\beta & 0 & s_\beta s_\theta  & s_\beta c_\theta  \\
0 & -c_\beta  & s_\beta c_\theta & -s_\beta s_\theta  \\
\end{array}
\right)\left(
\begin{array}{c}
A'_1 \\
A'_2\\
\varphi_A\\
\phi_A \\
\end{array}
 \right)
\eea Two  massless physical fields are $A_5$ and $A_6$.

\section{Mixing matrix for charged scalars}
\label{pl3}

\bea \left(
 \begin{array}{c}
\chi \\
 \chi' \\
\rho_1 \\
\rho_2\\
\rho'_1\\
\rho'_2 \\
\end{array}
\right) &=&  \left(
 \begin{array}{cccccc}
  \mathcal{M}_{11} & \mathcal{M}_{12}
& \mathcal{M}_{13}
 & \mathcal{M}_{14}& 0 & 0 \\
\mathcal{M}_{21} & \mathcal{M}_{22} & \mathcal{M}_{23}
 & \mathcal{M}_{24} & 0 & 0 \\
m_{31}s_\theta & m_{32}s_\theta & m_{33}s_\theta  & m_{34}s_\theta
& -c_\theta s_\gamma & -c_\theta c_\gamma \\
m_{31}c_\theta & m_{32}c_\theta  & m_{33}c_\theta & m_{34}c_\theta
 & s_\theta s_\gamma & s_\theta c_\gamma \\
m_{41}s_\theta & m_{42}s_\theta & m_{43}s_\theta & m_{44}s_\theta
& c_\theta c_\gamma & -c_\theta s_\gamma \\
m_{41}c_\theta & m_{42}c_\theta & m_{43}c_\theta & m_{44}c_\theta
& -s_\theta c_\gamma & s_\theta s_\gamma \\
\end{array}
\right)\left(
\begin{array}{c}
\zeta_1 \\
\zeta_2\\
\zeta_3\\
\zeta_4 \\
\varrho_1 \\
\varrho_2\\
\end{array}
 \right)
\eea where \bea \mathcal{M}_{11}&=&m_{11}s_\beta+m_{21}c_\beta, \
\mathcal{M}_{12}=m_{12}s_\beta+m_{22}c_\beta,\\
\mathcal{M}_{13}&=&m_{13}s_\beta+m_{23}c_\beta,\
\mathcal{M}_{14}= m_{14}s_\beta+m_{24}c_\beta,\\
\mathcal{M}_{21}&=&m_{21}s_\beta-m_{11}c_\beta, \
\mathcal{M}_{22} = m_{22}s_\beta-m_{12}c_\beta,\\
\mathcal{M}_{23}&= &m_{23}s_\beta-m_{13}c_\beta, \
\mathcal{M}_{24}= m_{24}s_\beta-m_{14}c_\beta
 \eea

 and
 \bea
m_{11}&=&\frac{g\sqrt{(t^2_\beta+1)(u^{'2}+w^{'2})}}{2\sqrt{m^2_{\zeta_4}}},\\
m_{12}&=&\frac{1}{\sqrt{u^{'2}+w^{'2}+v^2}}\frac{v}{\sqrt{1+t^2_\beta}},\\
m_{13}&=&\frac{1}{\sqrt{u^2+v^{'2}+w^2}}\frac{v't_\beta}{\sqrt{1
+t^2_\beta}},\\
m_{14}&=&\frac{k_1}{\sqrt{k^2_1+k^2_2+k^2_3+1}},\\
m_{21}&=&0=m_{33}=m_{42},\\
m_{22}&=&\frac{1}{\sqrt{u^{'2}+w^{'2}+v^2}}\frac{v'}{\sqrt{1+t^2_\beta}},\\
m_{23}&=&\frac{1}{\sqrt{u^2+v^{'2}+w^2}}\frac{-v'}{\sqrt{1
+t^2_\beta}},\\
m_{24}&=&\frac{k_2}{\sqrt{k^2_1+k^2_2+k^2_3+1}},\
m_{31}=\frac{-gv}{2\sqrt{m^2_{\zeta_4}}},\\
m_{32}&=&\frac{\sqrt{u^{'2}+w^{'2}}}{\sqrt{u^{'2}+w^{'2}+v^2}},\
m_{34}=\frac{k_3}{\sqrt{k^2_1+k^2_2+k^2_3+1}},\\
m_{41}&=&\frac{-gv'}{2\sqrt{m^2_{\zeta_4}}},\
m_{43}=\frac{\sqrt{u^{2}+w^{2}}}{\sqrt{v^{'2}+w^{2}+u^2}},\\
m_{44}&=&\frac{1}{\sqrt{k^2_1+k^2_2+k^2_3+1}} .\eea


\section{\label{pl4}Interactions of the $W$ boson with Higgs bosons}
\begin{table}
\bc \caption{\label{tab3}Trilinear couplings of  $W^+W^-$ with
neutral Higgs bosons.}

\vs

\begin{tabular}{|c|c|}\hline
Vertex  &   Coupling \\ \hline $W^+W^-S'_5$ &
$\frac{g^2vv'}{\sqrt{v^2+v^{'2}}}$ \\
  $W^+W^-\varphi_{s_{a36}}$ & $\frac{g^2c_\alpha}{2}\frac{v^{'2}-v^2}{\sqrt{v^{'2}+v^2}}$
  \\
  $W^+W^-\phi_{s_{a36}}$ & $-\frac{g^2s_\alpha}{2}\frac{v^{'2}-v^2}{\sqrt{v^{'2}+v^2}}$
  \\
\hline
\end{tabular}
\ec
\end{table}

\begin{table}
\caption{\label{tab4}Trilinear coupling constants of $W^-$ with
two Higgs bosons.}

\vs

\begin{tabular}{|c|c|}\hline
Vertex  &   Coupling \\ \hline  $W^{\mu -}\zeta^+_x
\stackrel{\leftrightarrow}{\pa_\mu}A'_1$ &
$-\fr{gw(2m_{2x}uu'+m_{1x}(u^2-u'^2))}{2(u^2+u'^2)\sqrt{u^2+w^2}}, \
x = 1,2,3,4.$\\  $W^{\mu -}\zeta^+_x
\stackrel{\leftrightarrow}{\pa_\mu}\varphi_A$ &
$-\fr{gu^2w(m_{2x}u-m_{1x}u')}{2(u^2+u'^2)(u^2+w^2)},\  x =
1,2,3,4.$\\  $W^{\mu -}\zeta^+_x
\stackrel{\leftrightarrow}{\pa_\mu}A'_2$ &
$\fr{gu(2m_{2x}uu'+m_{1x}(u^2-u'^2))}{2(u^2+u'^2)\sqrt{u^2+w^2}},\ x
= 1,2,3,4. $ \\  $W^{\mu -}\zeta^+_x
\stackrel{\leftrightarrow}{\pa_\mu} H^0_X$ &
$-\fr{ig(-2m_{1x}uu'+m_{2x}(u^2-u'^2))}{\sqrt{2}(u^2+u'^2)}, \ x =
1,2,3,4. $ \\  $W^{\mu -}
\zeta^+_x\stackrel{\leftrightarrow}{\pa_\mu}\varphi_{S_{24}} $ &
$\fr{ig(2m_{2x}uu'+m_{1x}(u^2-u'^2))}{2(u^2+u'^2)}, \ x = 1,2,3,4. $
\\ \hline
\end{tabular}
\end{table}

\begin{table}
\caption{\label{tab5}Nonzero quartic coupling constants of $W^+W^-$
with Higgs bosons.}

\vs

\begin{tabular}{|c|c|c|c|}\hline
Vertex  & Coupling  & Vertex & Coupling \\ \hline  $W^+W^- A'_1
A'_2$ & $-\fr{g^2uw}{2(u^2+w^2)}$ &
 $W^+W^-\varphi_{S_{a36}} \phi_{S_{a36}} $ & $ -\fr{g^2s_{2\alpha}}{4}
$\\  $W^+W^-  A'_1 A'_1$ & $
 \fr{g^2w^2}{2(u^2+w^2)}$ &
$W^+W^-  A'_2 A'_2 $ &  $ \fr{g^2u^2}{2(u^2+w^2)} $
\\
$W^+ W^-  A_5 A_5$ & $ \fr{g^2}{2} $ &
 $W^+W^- \varphi_{S_{a36}}\varphi_{S_{a36}}$ & $
\fr{g^2c^2_\alpha}{2} $ \\  $W^+W^-\phi_{S_{a36}} \phi_{S_{a36}} $ &
$ \fr{g^2s^2_\alpha}{2}$& $W^+W^- \zeta^+_x\zeta^-_y $ & $
\fr{g^2(m_{1x}m_{1y}+m_{2x}m_{2y})}{2},$\\&&&$ x,y = 1,2,3,4. $
\\ \hline
\end{tabular}
\end{table}

\section{\label{pl5}Interactions of the $Z$ boson with Higgs bosons}
\begin{table}
\caption{\label{tab6}Trilinear coupling constants of $Z^\mu$ with
two charged Higgs bosons.}

\vs

\begin{tabular}{|c|c|}\hline
Vertex  &   Coupling \\ \hline  $Z^\mu \varrho^-_1
\stackrel{\leftrightarrow}{\pa_\mu} \varrho^+_1 $ &
$-\fr{ig}{6(u^2+w^2)}
\left\{-2\sqrt{3}u^2(U_{22}-2s_W\sqrt{\fr{1}{3-4s^2_W}}U_{32})
-6uU_{42}w+\right.$\\&$\left.
[3U_{12}+\sqrt{3}(U_{22}+4s_W\sqrt{\fr{1}{3-4s^2_W}}U_{32})]w^2\right\}
 $\\
$Z^\mu  \zeta^-_x\stackrel{\leftrightarrow}{\pa_\mu} \varrho^+_1 $ &
$-\fr{ig}{2\sqrt{v^2+v'^2}(u^2+w^2)}
(m_{4x}v-m_{3x}v')\left[-u^2U_{42}+ \right.$\\&$\left.
u(U_{12}+\sqrt{3}U_{22})w+U_{42}w^2\right], \ x=1,2,3,4.$\\  
 $Z^\mu  \zeta^-_x\stackrel{\leftrightarrow}{\pa_\mu}
\varrho^+_2 $ & $\fr{ig}{2\sqrt{v^2+v'^2}(u^2+w^2)}
(m_{3x}v+m_{4x}v')\left[-u^2U_{42}+\right.$\\&$\left.
u(U_{12}+\sqrt{3}U_{22})w+U_{42}w^2\right] , \ x=1,2,3,4.$\\
$Z^\mu\zeta^-_x\stackrel{\leftrightarrow}{\pa_\mu}\zeta^+_y  $ & $
-\fr{ig}{6s_W(u^2+w^2)}\left\{(m_{1y}m_{1x}+m_{2y}m_{2x})
[3U_{12}-\sqrt{3}(U_{22}-\right.$\\&$
2s_W\sqrt{\fr{1}{3-4s^2_W}}U_{32})](u^2+w^2)
+(m_{3y}m_{3x}+m_{4y}m_{4x})(u^2[3U_{12}+$\\&$
\sqrt{3}(U_{22}+4s_W\sqrt{\fr{1}{3-4s^2_W}}U_{32})]+$\\&$\left.
6uwU_{42}-2\sqrt{3}w^2(U_{22}-2s_W\sqrt{\fr{1}{3-4s^2_W}}U_{32}))\right\},
 \ x,y=1,2,3,4.$\\ \hline
\end{tabular}
\end{table}

\begin{table}
\caption{\label{tab7}Trilinear couplings of $Z_\mu$ with two neutral
Higgs bosons.}

\vs

\begin{tabular}{|c|c|c|c|}\hline
Vertex  & Coupling  & Vertex & Coupling \\ \hline $Z^\mu  A'_1
\stackrel{\leftrightarrow}{\pa_\mu} S'_{1a} $ & $-\fr{g
uu'}{3(u^2+u'^2)\sqrt{u^2+w^2}}f_1 $& $Z^\mu
A'_2\stackrel{\leftrightarrow}{\pa_\mu} S'_{1a} $ & $-\fr{g
uu'}{3(u^2+u'^2)\sqrt{u^2+w^2}}f_7
 $\\
$Z^\mu  \varphi_A\stackrel{\leftrightarrow}{\pa_\mu}  S'_{1a}  $
 & $ \fr{g(u^2-u'^2)}{6(u^2+u'^2)(u^2+w^2)}f_8 $
& $Z^\mu A_5 \stackrel{\leftrightarrow}{\pa_\mu}
\varphi_{S_{a36}}$
 & $ -\fr{gvc_\alpha}{6\sqrt{v^2+v'^2}} f_2
$\\
$Z^\mu A_6 \stackrel{\leftrightarrow}{\pa_\mu} \varphi_{S_{a36}} $
 & $ -\fr{gv'c_\alpha}{6\sqrt{v^2+v'^2}} f_2$&
 $Z^\mu  A'_2 \stackrel{\leftrightarrow}{\pa_\mu} \varphi_{S_{a36}}
$ & $\fr{g s_\alpha(u^2-u'^2)}{6(u^2+u'^2)\sqrt{u^2+w^2}}f_7 $\\
$Z^\mu  A'_1 \stackrel{\leftrightarrow}{\pa_\mu} \varphi_{S_{a36}}
$ & $\fr{g s_\alpha(u^2-u'^2)}{6(u^2+u'^2)\sqrt{u^2+w^2}}f_1 $&
$Z^\mu  \varphi_A\stackrel{\leftrightarrow}{\pa_\mu}
\varphi_{S_{a36}}  $
 & $ \fr{gs_\alpha uu'}{3(u^2+u'^2)(u^2+w^2)} f_8
$\\
$Z^\mu A_5\stackrel{\leftrightarrow}{\pa_\mu}S'_5  $
 & $ \fr{gv'}{6\sqrt{v^2+v'^2}} f_2$&
$Z^\mu A_6\stackrel{\leftrightarrow}{\pa_\mu}  S'_5  $
 & $ -\fr{gv}{6\sqrt{v^2+v'^2}} f_2
$ \\ $Z^\mu A_5 \stackrel{\leftrightarrow}{\pa_\mu} \phi_{S_{a36}} $
 & $ \fr{gvs_\alpha}{6\sqrt{v^2+v'^2}} f_2$&
$Z^\mu A_6 \stackrel{\leftrightarrow}{\pa_\mu} \phi_{S_{a36}} $
 & $ \fr{gv's_\alpha}{6\sqrt{v^2+v'^2}} f_2
$\\ $Z^\mu  A'_1 \stackrel{\leftrightarrow}{\pa_\mu} \phi_{S_{a36}}
$ & $\fr{g c_\alpha(u^2-u'^2)}{6(u^2+u'^2)\sqrt{u^2+w^2}} f_1 $&
$Z^\mu  A'_2 \stackrel{\leftrightarrow}{\pa_\mu} \phi_{S_{a36}} $ &
$\fr{g c_\alpha(u^2-u'^2)}{6(u^2+u'^2)\sqrt{u^2+w^2}} f_7
 $\\   $Z^\mu A'_1\stackrel{\leftrightarrow}{\pa_\mu} \varphi_{S_{24}}  $
 & $ \fr{g }{6\sqrt{u^2+w^2}}\left[-3uU_{42}+(3U_{12}+\right.$\\&$
 \left.\sqrt{3}(U_{22}-2s_W\sqrt{\fr{1}{3-4s^2_W}}
U_{32}))w\right] $& $Z^\mu
\varphi_A\stackrel{\leftrightarrow}{\pa_\mu} \phi_{S_{a36}}  $
 & $ \fr{gc_\alpha uu'}{3(u^2+u'^2)(u^2+w^2)} f_8 $\\
$Z^\mu A'_2\stackrel{\leftrightarrow}{\pa_\mu}\varphi_{S_{24}}   $ &
$\fr{g }{6\sqrt{u^2+w^2}}\left[2\sqrt{3}u(U_{22}+\right.$\\&$
 \left.s_W\sqrt{\fr{1}{3-4s^2_W}}
U_{32})+3w U_{42}\right]
 $\\
  \hline
\end{tabular}
\end{table}

\begin{table}
\caption{\label{tab8a}Trilinear coupling constants of $ZZ$ with one
scalar bosons.}

\vs

\begin{tabular}{|c|c|}\hline
Vertex  & Coupling \\ \hline $Z Z S'_5$ & $
\fr{g^2vv'}{3\sqrt{v^2+v'^2}}f_4$\\
$Z Z S'_{1a}$ & $
\fr{g^2u'}{3\sqrt{u^2+u'^2}\sqrt{u^2+w^2}}(u^2f_5+2uwU_{42}+w^2f_6)$\\
 $Z Z \varphi_{S_{24}}$ & $ \fr{g^2\sqrt{u^2+u'^2}}{6u\sqrt{u^2+w^2}}
 [(u^2-w^2)U_{42}f_2+uwf_3]$\\
 $Z Z \varphi_{S_{a36}}$ & $ \frac{g^2}{4}
 \left\{[-(a_1s_\theta+a_2c_\theta)s_\beta+(a_3s_\theta+a_4c_\theta)c_\beta]s_\alpha
+\frac{v'-v}{\sqrt{v^{2}+v{'}^{2}}}c_\alpha\right\}
 $\\
 $Z Z \phi_{S_{a36}}$ & $ \frac{g^2}{4}
 \left\{[-(a_1s_\theta+a_2c_\theta)s_\beta+(a_3s_\theta+a_4c_\theta)c_\beta]c_\alpha
-\frac{v'-v}{\sqrt{v^{2}+v{'}^{2}}}s_\alpha\right\}
  $\\ \hline
\end{tabular}
\end{table}

\begin{table}
\caption{\label{tab9}Non-zero quartic coupling constants of $ZZ$
with two neutral scalar  bosons.}

\vs

\begin{tabular}{|c|c|}\hline
Vertex  & Coupling \\ \hline $ZZ A'_1 A'_2$&$\fr{g^2}{6}U_{42}f_2 $
\\
 $Z Z A'_1 A'_1$ & $ \fr{g^2}{6}f_5 $
\\
$Z Z A'_2 A'_2$ & $ \fr{g^2}{6}f_6  $
\\
$Z Z H H $ & $ \fr{g^2}{6}f_4,\ H=S'_5, A_5, A_6
 $\\
$ZZ S'_{1a}\varphi_{S_{24}}$ &
$\fr{g^2uu'}{3(u^2+u'^2)+(u^2+w^2)}[(u^2-w^2)U_{42}f_2+uwf_3] $ \\
$ZZ \varphi_{S_{24}} \phi_{S_{a36}}$ &
$-\fr{g^2c_\alpha(u^2-u'^2)}{6(u^2+u'^2)+(u^2+w^2)}[(u^2-w^2)U_{42}f_2+uwf_3]
$ \\ $ZZ \varphi_{S_{24}} \varphi_{S_{a36}}$ &
$-\fr{g^2s_\alpha(u^2-u'^2)}{6(u^2+u'^2)+(u^2+w^2)}[(u^2-w^2)U_{42}f_2+uwf_3]
$ \\
$Z Z S'_{1a} S'_{1a}$ & $
\fr{g^2}{6(u^2+w^2)}(u^2f_5+2uwU_{42}f_2+w^2f_6)$\\
$Z Z \varphi_{S_{24}} \varphi_{S_{24}}$ & $
 \fr{g^2}{6(u^2+w^2)}(u^2f_6-2uwU_{42}f_2+w^2f_5) $
 \\ $Z Z H^0_X H^{0*}_X $ & $
 \fr{g^2}{6(u^2+w^2)}(u^2f_6-2uwU_{42}f_2+w^2f_5) $
 \\$Z Z \varphi_{S_{a36}}\varphi_{S_{a36}}$ & $
\fr{g^2}{6(u^2+w^2)}[c^2_\alpha(u^2+w^2)f_4+
s^2_\alpha(u^2f_5+2uwU_{42}f_2+w^2f_6)] $
\\$Z Z \varphi_A \varphi_A $ & $
\fr{g^2}{6(u^2+w^2)}(u^2f_5+2uwU_{42}f_2+w^2f_6) $
\\$Z Z\phi_{S_{a36}}\phi_{S_{a36}}$ & $
\fr{g^2}{6(u^2+w^2)}[s^2_\alpha(u^2+w^2)f_4+
c^2_\alpha(u^2f_5+2uwU_{42}f_5+w^2f_6)] $
\\$ZZ \varphi_{S_{a36}} \phi_{S_{a36}}$ &
$\fr{g^2s_{2\alpha}}{12(u^2+w^2)}\left\{u^2\left[4\sqrt{3}U_{12}(U_{22}+
s_W\sqrt{\fr{1}{3-4s^2_W}}U_{32})+3U^2_{42}-\right.\right.
$\\&$\left.\fr{12s_WU_{22}U_{32}}{\sqrt{3-4s^2_W}}+
\fr{12s^2_WU^2_{32}}{-3+4s^2_W}\right]
-2uwU_{42}\left[-3U_{12}+\right.$\\&$\left.\sqrt{3}(U_{22}+
4s_W\sqrt{\fr{1}{3-4s^2_W}}U_{32})\right]+\left[-3U^2_{12}+
2\sqrt{3}U_{12}\left(U_{22}+\right.\right.$\\&
$\left.\left.\left.4s_W\sqrt{\fr{1}{3-4s^2_W}}
U_{32}\right)+3(U^2_{22}+U^2_{42}+\fr{4s^2_W
U^2_{32}}{-3+4s^2_W})\right]w^2 \right\} $ \\  \hline
\end{tabular}
\end{table}

\begin{table}
\caption{\label{tab11}Non-zero quartic coupling constants of $ZZ$
with two charged bosons.}

\vs

\begin{tabular}{|c|c|}\hline
Vertex  & Coupling \\ \hline $Z Z \varrho^-_1 \varrho^+_1$ & $
\fr{g^2}{6(u^2+w^2)}\left\{u^2[4U^2_{22}-\fr{16s_WU_{22}U_{32}}{\sqrt{3-4s^2_W}}-
\fr{16s^2_WU^2_{32}}{-3+4s^2_W}+3U^2_{42}]-\right.$\\&$\left.
2uw[3U_{12}-\sqrt{3}(U_{22}-\fr{8s_WU_{32}}{\sqrt{3-4s^2_W}})]U_{42}
+w^2\left[3U^2_{12}+U^2_{22}+\fr{8s_WU_{22}U_{32}}{\sqrt{3-4s^2_W}}-
\right.\right.$\\&$\left.\left.\fr{16s^2_WU^2_{32}}{-3+4s^2_W}+
2\sqrt{3}U_{12}(U_{22}+\fr{4s_WU_{32}}{\sqrt{3-4s^2_W}})+3U^2_{42}\right]\right\}$
\\
$Z Z \varrho^-_1 \zeta^+_x$ & $
-\fr{g^2(m_{4x}v-m_{3x}v')}{6(u^2+w^2)\sqrt{v^2+v'^2}}
\left\{(u^2-w^2)[3U_{12}-\sqrt{3}(U_{22}-\fr{8s_WU_{32}}{\sqrt{3-4s^2_W}})]U_{42}-
\right.$\\&$\left.uw[3U^2_{12}-3U_{22}(U_{22}-\fr{8s_WU_{32}}{\sqrt{3-4s^2_W}})+
2\sqrt{3}U_{12}(U_{22}+\fr{4s_WU_{32}}{\sqrt{3-4s^2_W}})]\right\}
,\ x=1,2,3,4 $
\\
$Z Z \varrho^-_2 \zeta^+_x$ & $
\fr{g^2(m_{3x}v+m_{4x}v')}{6(u^2+w^2)\sqrt{v^2+v'^2}}
\left\{(u^2-w^2)[3U_{12}-\sqrt{3}(U_{22}-\fr{8s_WU_{32}}{\sqrt{3-4s^2_W}})]U_{42}-
\right.$\\&$\left.uw[3U^2_{12}-3U_{22}(U_{22}-\fr{8s_WU_{32}}{\sqrt{3-4s^2_W}})+
2\sqrt{3}U_{12}(U_{22}+\fr{4s_WU_{32}}{\sqrt{3-4s^2_W}})]\right\}
,\ x=1,2,3,4$
\\
$Z Z \zeta^-_y \zeta^+_x$ & $ \fr{g^2}{6(u^2+w^2)}
\left\{(m_{1y}m_{1x}+m_{2y}m_{2x})(u^2+w^2)\left[3U^2_{12}+
U^2_{22}-\right.\right.$\\&$\left.\fr{4s_WU_{22}U_{32}}{\sqrt{3-4s^2_W}}-
\fr{4s^2_WU^2_{32}}{-3+4s^2_W}-2\sqrt{3}U_{12}(U_{22}
-\fr{2s_WU_{32}}{\sqrt{3-4s^2_W}})\right]+
$\\&$(m_{3y}m_{3x}+m_{4y}m_{4x})[u^2\left(3U^2_{12}+
U^2_{22}+\fr{8s_WU_{22}U_{32}}{\sqrt{3-4s^2_W}}-
\right.$\\&$\left.\fr{16s^2_WU^2_{32}}{-3+4s^2_W}+2\sqrt{3}U_{12}(U_{22}+
\fr{4s_WU_{32}}{\sqrt{3-4s^2_W}})+3U^2_{42}\right)+2uw(3U_{12}-
\sqrt{3}\left(U_{22}-\right.$\\&$\left.\left.\fr{8s_WU_{32}}{\sqrt{3-4s^2_W}}\right))U_{42}+
w^2(4U^2_{22}-\fr{16s_WU_{22}U_{32}}{\sqrt{3-4s^2_W}}-
\fr{16s^2_WU^2_{32}}{-3+4s^2_W}+3U^2_{42})]\right\}$
\\ \hline
\end{tabular}
\end{table}

\section{Trilinear interactions of the $W, Z$ boson with Higgs bosons }
\label{pl6}

\begin{table}
\caption{\label{tab13}Trilinear coupling constants of $W^+$ with
neutral gauge boson and the charged scalar boson.}

\vs

\begin{tabular}{|c|c|}\hline
Vertex  & Coupling \\ \hline  $Z W^+ \varrho^-_2$ & $\fr{g^2
\sqrt{v^2+v'^2}}{12(u^2+w^2)}\left\{u^2[3U_{12}+\sqrt{3}(U_{22}-
\fr{8s_WU_{32}}{\sqrt{3-4s^2_W}})]+6uwU_{42}-\right.$\\&$\left.
2\sqrt{3}w^2(U_{22}+\fr{4s_WU_{32}}{\sqrt{3-4s^2_W}})\right\} $\\
$Z W^+ \zeta^-_x$ & $-\fr{g^2
}{4u\sqrt{u^2+u'^2}(u^2+w^2)^\fr{3}{2}}[(u^2-w^2)U_{42}-uw(U_{12}+\sqrt{3}U_{22})]
\times$\\&$[m_{1x}(u^2-u'^2)(u^2+w^2)+
u(\sqrt{u^2+u'^2}\sqrt{u^2+w^2}(m_{3x}v+m_{4x}v')+$\\&$2m_{2x}u'(u^2+w^2))],\
x=1,2,3,4$
\\
$A W^+ \varrho^-_2$ & $\fr{g^2
\sqrt{v^2+v'^2}}{12(u^2+w^2)}\left\{u^2[3U_{11}+
\sqrt{3}(U_{21}-\fr{8s_WU_{31}}{\sqrt{3-4s^2_W}})]+
6uwU_{41}-\right.$\\&$\left.2\sqrt{3}w^2(U_{21}+
\fr{4s_WU_{31}}{\sqrt{3-4s^2_W}})\right\}$\\
$A W^+ \zeta^-_x$ & $-\fr{g^2
}{4u\sqrt{u^2+u'^2}(u^2+w^2)^\fr{3}{2}}[(u^2-w^2)U_{41}-
uw(U_{11}+\sqrt{3}U_{21})] \times$\\&$[m_{1x}(u^2-u'^2)(u^2+w^2)+
u(\sqrt{u^2+u'^2}\sqrt{u^2+w^2}(m_{3x}v+m_{4x}v')+
$\\&$2m_{2x}u'(u^2+w^2))],\  x=1,2,3,4$
\\ \hline
\end{tabular}
\end{table}


\begin{thebibliography}{999}

\bibitem{superk} SuperKamiokande Collaboration, Y. Fukuda, et
al., Phys. Rev. Lett. 81 (1998) 1158; 81 (1998) 1562; 82 (1999)
2644; 85 (2000) 3999; Y. Suzuki, Nucl. Phys. B (Proc. Suppl.) 77
(1999) 35; S. Fukuda, et al., Phys. Rev. Lett. 86 (2001) 5651; Y.
Ashie, et al., Phys. Rev. Lett. 93 (2004) 101801; Phys. Rev. Lett.
97 (2006) 171801.

\bibitem{kam} KamLAND Collaboration, K. Eguchi, et al.,  Phys.
Rev. Lett. 90 (2003) 021802; T. Araki, et al., Phys. Rev. Lett. 94
(2005) 081801.

\bibitem{sno} SNO Collaboration, Q.R. Ahmad, et al., Phys.
Rev. Lett. 89 (2002) 011301; 89 (2002) 011302; 92 (2004) 181301; B.
Aharmim, et al., Phys. Rev. C 72 (2005) 055502.

\bibitem{susy} J. Wess, J. Bagger, Supersymmetry and Supergravity, second ed., Princeton Univ. Press,
Princeton, NJ, 1992; H.E. Haber, G.L. Kane, Phys. Rep. 117 (1985)
75.

\bibitem{ppf} F. Pisano, V. Pleitez, Phys. Rev. D 46 (1992) 410;
P.H. Frampton, Phys. Rev. Lett. 69 (1992) 2889; R. Foot, et al.,
Phys. Rev. D 47 (1993) 4158.

\bibitem{flt} M. Singer, et al., Phys. Rev. D 22 (1980) 738.

\bibitem{331rh} R. Foot, H.N. Long, T.A. Tran, Phys. Rev. D 50 (1994) 34R;
J.C. Montero, et al., Phys. Rev. D 47 (1993) 2918; H.N. Long, Phys.
Rev. D 54 (1996) 4691; 53 (1996) 437.

\bibitem{chargequan} C.A. de S. Pires, O.P. Ravinez,
Phys. Rev. D 58 (1998) 035008; A. Doff, F. Pisano,  Mod. Phys. Lett.
A 14 (1999) 1133; Phys. Rev. D 63 (2001) 097903; P.V. Dong, H.N.
Long,  Int. J. Mod. Phys. A 21 (2006) 6677.

\bibitem{ponce}  W.A. Ponce, Y. Giraldo, L.A. Sanchez, Phys. Rev. D 67 (2003) 075001.

\bibitem{haihiggs} P.V. Dong, H.N. Long, D.T. Nhung, D.V. Soa, Phys. Rev. D 73 (2006) 035004.

\bibitem{higgseconom} P.V. Dong, H.N. Long, D.V. Soa, Phys. Rev. D 73 (2006) 075005.

\bibitem{dlhh} P.V. Dong, T.T. Huong, D.T. Huong, H.N. Long, Phys. Rev. D 74 (2006) 053003.

\bibitem{dls1} P.V. Dong, H.N. Long, D.V. Soa, Phys. Rev. D 75 (2007) 073006.

\bibitem{susyec} P.V. Dong, D.T. Huong, M.C. Rodriguez, H. N. Long, Nucl. Phys. B 772 (2007) 150.

\bibitem{logan} M. Duhrssen, et al., Phys. Rev. D 70 (2004) 113009.

\bibitem{changlong} D. Chang, H.N. Long, Phys. Rev. D 73 (2006) 053006.

\bibitem{scalrh} M.B. Tully, G.C. Joshi, Phys. Rev. D 64 (2001) 011301R;
R.A. Diaz, R. Martinez, F. Ochoa, Phys. Rev. D 69 (2004) 095009.

\bibitem{hrl} D.T. Huong, M.C. Rodriguez, H.N. Long, Scalar sector of supersymmetric $\mbox{SU}(3)_C\otimes
\mbox{SU}(3)_L \otimes \mbox{U}(1)_N$ model with right-handed
neutrinos, hep-ph/0508045.

\bibitem{shiggs} J.C. Montero, V. Pleitez, M.C. Rodriguez, Phys. Rev. D 65 (2002) 035006;
M.C. Rodriguez, Int. J. Mod.  Phys. A 21 (2006) 4303.

\bibitem{kame} E. Asakawa, S. Kanemura, Phys. Lett. B 626 (2005) 111.

\bibitem{roy} See for example, D.P. Roy, hep-ph/0510070.

\bibitem{s331r} J.C. Montero, V. Pleitez, M.C. Rodriguez, Phys. Rev. D 70 (2004) 075004.

\bibitem{dl} P.V. Dong, H.N. Long, Eur. Phys. J. C 42 (2005) 325.

\bibitem{pdg} W.-M. Yao, et al., Particle Data Group,
J. Phys. G: Nucl. Part. Phys. 33 (1) (2006) 32.

\bibitem{mal} J.E.C. Montalvo, M.D. Tonasse, Phys. Rev. D 71 (2005) 095015.

\bibitem{ninhlong} L.D. Ninh, H.N. Long, Phys. Rev. D 72 (2005) 075004.

\bibitem{longinami}H.N. Long and T. Inami, Phys. Rev. D 61 (2000)
075002.

\bibitem{tullyjoshi} M.B. Tully, G.C. Joshi, Phys. Lett. B 466 (1999) 333.

\bibitem{glob} J.F. Gunion, et al., The Higgs Hunter's Guide, Addison-Wesley, New York, 1990.

\bibitem{kane} G. Kane, W. Repko, W. Rolnick, Phys. Lett. B 148 (1984) 367;
M. Chanowiz, M.K. Gaillard, Phys. Lett. B 142 (1984) 85; S. Dawson,
Nucl. Phys. B 249 (1985) 42.

\bibitem{cteq6}  http://user.pa.msu.edu/wkt/cteq/cteq6/cteq6pdf.html

\end{thebibliography}
\end{document}